\documentclass[usenatbib]{mn2e}
\usepackage{fixltx2e}     %preserve figure order when 1/2-column figures mixed
\usepackage{epsfig}
\usepackage{natbib}		% bibliography
\newcommand{\Myr}{\,\mbox{Myr}}

\newcommand{\Mpc}{\,\mbox{Mpc}}
\newcommand{\kpc}{\,\mbox{kpc}}
\newcommand{\kms}{\,\mbox{km}\,\mbox{s}^{-1}}

\newcommand{\msun}{\,M_{\sun}}
\newcommand{\degree}{^\circ}
\newcommand{\amin}{\,\mbox{arcmin}}

\newcommand{\mtwelve}{\times 10^{12}\msun}
\newcommand{\tsim}{\sim\!}

\newcommand{\nbody}{$N$-body}
\newcommand{\feh}{[\mbox{Fe}/\mbox{H}]}
\newcommand{\PA}{\mathit{PA}}
\newcommand{\pos}{\mathit{pos}}
\newcommand{\im}{\mathit{im}}
\newcommand{\gtrsim}{\ga}
\newcommand{\ltrsim}{\la}

\defcitealias{fardal07}{F07}
\title[Inferring Andromeda's Mass from its Giant Stream] %mnras
  {Inferring the Andromeda Galaxy's Mass from its Giant Southern Stream with Bayesian Simulation Sampling}
\author[M. A. Fardal et al.]       %mnras
{Mark A. Fardal$^1$\thanks{E-mail: fardal@astro.umass.edu}, 
Martin D. Weinberg$^1$, 
Arif Babul$^2$,
Mike J. Irwin$^3$, \newauthor
Puragra Guhathakurta$^4$, 
Karoline M. Gilbert$^{5,11}$, 
Annette M. N. Ferguson$^6$, \newauthor
Rodrigo A. Ibata$^7$,
Geraint F. Lewis$^8$,
Nial R. Tanvir$^9$,
Avon P. Huxor$^{10}$\\
$^1$Dept.\ of Astronomy, University of Massachusetts, 
    Amherst, MA 01003, USA \\
$^2$Dept.\ of Physics \& Astronomy, University of Victoria, 
    Elliott Building, 3800 Finnerty Rd., Victoria, BC, V8P 1A1, Canada \\
$^3$Institute of Astronomy, University of Cambridge, Madingley Road, Cambridge CB3 0HA \\
$^4$UCO/Lick Observatory, Dept.\ of Astronomy \& Astrophysics,
     Univ. of California, 1156 High St., Santa Cruz, CA 95064, USA \\
$^5$Department of Astronomy, University of Washington, Box 351580, 
    Seattle, WA 98195-1580, USA \\
$^6$Institute for Astronomy, University of Edinburgh, Royal Observatory, 
    Blackford Hill, Edinburgh EH9 3HJ \\
$^7$Observatoire Astronomique, Universit\'{e} de Strasbourg, CNRS, 11 rue de l'Universit\'{e}, 
    F-67000 Strasbourg, France \\
$^8$Sydney Institute for Astronomy, School of Physics, A28, 
    University of Sydney, NSW 2006, Australia \\
$^9$Department of Physics and Astronomy, University of Leicester, 
    University Road, Leicester LE1 7RH \\
$^{10}$Astronomisches Rechen-Institut, Universit\"{a}t Heidelberg, M\"{o}nchhofstrasse 12-14, 
    69120 Heidelberg, Germany \\
 $^{11}$Hubble Fellow
}  %don't put a \\ after last author
\date{Submitted to MNRAS \today}
\pagerange{\pageref{firstpage}--\pageref{lastpage}} \pubyear{2012} %mnras
\begin{document}
\maketitle  %mnras
\label{firstpage}
\begin{abstract}
M31 has a giant stream of stars extending far to the south and a great
deal of other tidal debris in its halo, much of which is thought to be
directly associated with the southern stream.  We model this structure
by means of Bayesian sampling of parameter space, where each sample
uses an $N$-body simulation of a satellite disrupting in M31's
potential.  We combine constraints on stellar surface densities from
the Isaac Newton Telescope survey of M31 with kinematic data and
photometric distances.  This combination of data tightly constrains
the model, indicating a stellar mass at last pericentric passage of
$\log_{10} \, (M_s /\msun) = 9.5 \pm 0.1$, comparable to the LMC.  Any
existing remnant of the satellite is expected to lie in the NE Shelf
region beside M31's disk, at velocities more negative than M31's disk
in this region.  This rules out the prominent satellites M32 or
NGC~205 as the progenitor, but an overdensity recently discovered in
M31's NE disk sits at the edge of the progenitor locations found in
the model.  M31's virial mass is constrained in this model to be
$\log_{10} \, M_{200} = 12.3 \pm 0.1$, alleviating the previous
tension between observational virial mass estimates and expectations
from the general galactic population and the timing argument.  The
techniques used in this paper, which should be more generally
applicable, are a powerful method of extracting physical inferences
from observational data on tidal debris structures.
\end{abstract}

\section{INTRODUCTION} %1
\label{sec.intro}
In the current model of structure formation, the stellar halos of
galaxies are thought to be messy, inhomogeneous places, populated not
only by satellite galaxies and globular clusters but also by tidal
debris structures that are mixed to varying degrees
\citep{searle78, bullock05,abadi06, johnston08, cooper10}. 
Observations give strong support to this picture: for example, the MW
contains numerous streams of stars ripped from dwarf satellites and
globulars by the galactic potential.  These range from the Sagittarius
galaxy's tidal stream to numerous smaller streams, most of which have
no identified progenitor.  Our external view of M31 has enabled
photometric surveys that give us more complete (though shallower)
coverage than is available in the Milky Way. The inhomogeneities there
are even more impressive: a 150-kpc-long tidal stream to the south
\citep[the giant southern stream or GSS, ][]{ibata01,ibata07},
several tangential streams across the minor axis
\citep{ibata07,chapman08,alan09,tanaka10}, 
and numerous structures resembling blobs, spurs, or
shelves more than streams
\citep{ferguson02,ibata05,fardal07,ibata07,alan09}.  
Surveys of more distant galaxies have found additional tidal streams 
\citep[e.g.,][]{davidmd08, davidmd09}, showing that streams are 
common features of galaxy halos.

These cold structures are important in their own right: they are the
products of hierarchical galaxy formation, and studying the
information encoded in their kinematics and chemical composition can
lead to insight into the way structure forms on halo mass scales of
$\tsim 10^{12} \msun$ down to $\tsim 10^8 \msun$
\citep{bullock05, johnston08, font08, gilbert09lcdm}.  They are also
significant as kinematic probes of the halo mass, both in the Milky
Way \citep{helmi04, johnston05, grillmair09, willett09, koposov10} and
in M31 \citep{ibata04, fardal06}, though results from this approach
are still controversial.  While in other galaxies the mass can be
estimated by means of satellite tracers \citep{brainerd03,more09},
abundance at a given stellar mass 
\citep[e.g.][]{yang09,guo10,behroozi10}, or gravitational lensing
\citep{mandelbaum06}, these estimates are generally statistical in
nature, and thus may gloss over systematic differences between halos
of different galaxy types.  A well-determined mass profile using cold
kinematic tracers in the halo of the MW or in M31 would provide an
important test of these techniques.  Cold streams could also reveal
more subtle aspects such as halo prolateness/oblateness
\citep{helmi04, johnston05}, triaxiality \citep{law09}, or
substructure \citep{johnston02,carlberg09}.
Furthermore, the mass of M31's halo is important for
understanding this much-studied galaxy itself, 
as well as its satellite system, which is fundamental 
to the study of galaxy formation on small scales.

M31's giant southern stream (GSS) provides one of the best prospects
for such a mass determination, because it spans a large range in
radius and because its large stellar mass yields many spectroscopic
targets.  (Various arguments assign its initial stellar mass to be in
the range $10^8$ to $5 \times 10^9 \msun$, according to
\citealp{font06}.)  The stream falls in from behind M31 \citep{alan03}
and speeds up as it goes \citep{ibata04, raja06, kalirai06a,gilbert09}.

\citet{ibata04} and \citet{font06} attempted to fit progenitor orbits
to the stream to determine its orbit and constrain M31's halo mass.
However, there are two major problems with this method.  The first is
that the stream does not coincide with the orbit, as expected on
general grounds and shown explicitly with $N$-body simulations
\citep{fardal06}.  Especially for a highly radial, massive stream such
as the GSS, the trailing part of the stream consists of stars boosted
to much higher energies than that of the stream progenitor,
increasingly so as one goes further out in the stream.  As a partial
fix, \citet{fardal06} found an approximation relating the progenitor
orbit to the stream (extending work by \citealp{johnston98}), enabling
the central path of the stream through phase space to be estimated
without the use of $N$-body simulations.  Even so, the wide range of
possible energy boosts rendered estimates of the halo mass and orbit
highly uncertain, and \citet{fardal06} could only state that any
continuation of the stream must be to the NE of M31's center.

The second problem is that an approximation based on the central path
of the stream cannot make effective use of the full debris
distribution, especially that of subsequent wraps which may appear as
widely dispersed tidal features.  In 
\citet[][henceforth F07]{fardal07}, we used $N$-body simulations to
argue that red giant branch (RGB) stars and planetary nebulae (PNe)
seen in and around the NE half of M31's disk constituted a second wrap
of the orbit, while a faint structure seen to the W side of the disk
was a third wrap (see Figure~\ref{fig.map}).  This remarkably
successful scenario has received subsequent support from studies of
the stellar population in these regions \citep{richardson08}, from
spectroscopic discovery of the apparent fourth wrap of the orbit
\citep{gilbert07}, and from measurements of the velocity distribution
in the W Shelf region \citep{fardal12}.  The model can be refined by
adding rotation to the progenitor, which improves the transverse
profile of the stream \citep{fardal08}.

In this paper we seek to improve on the model of F07 by quantitative
comparison of simulations and observations.  Using simulations to
improve and constrain the model is difficult, however.  Simulations
are slow, and in addition intrinsically stochastic, which presents
difficulties for gradient-based fitting techniques and uncertainties
based on second derivatives of the likelihood.  Therefore, we have
chosen to explore parameter space with sampling techniques commonly
used in Bayesian statistics.  Our approach is related to several prior
efforts to automate fitting of merger debris using methods such as
genetic algorithms \citep{theis01, howley08}.  However, these efforts
have only sought a best-fit model of merger debris without quantifying
the parameter uncertainties.  Our goal here is to 
{\em sample the full parameter distribution}, in order to draw
scientifically useful conclusions (i.e., with uncertainty estimates)
about interesting issues such as the mass of the progenitor and of
M31's halo.  The combination of Bayesian sampling with $N$-body
simulations of specific structures has not been tried before to our
knowledge, and one might expect a very large number of $N$-body
simulations would be required.  This paper therefore serves among
other purposes to test whether the method is currently feasible.

The rest of the paper starts by describing in Section~\ref{sec.methods} 
our methods for sampling from parameter space.  We begin with a sketch
of the procedure before plunging into the details.  We then describe
the physical collision model and its parameterization, the likelihood
function which incorporates constraints from star-count maps as well
as velocity and distance measurements along the GSS, and the method
used to sample from parameter space.  Section~\ref{sec.results}
presents our results, in terms of both the distributions of the model
parameters and of the GSS debris structures generated by the
simulation samples.  We find tight constraints on most of our
parameters, including M31's halo mass, and significant correlations
among some of them.  We illustrate both the characteristic features
and the variable aspects of the simulations, test the goodness of fit
of the models, check their validity using observables not included in
the likelihood function, and offer predictions for future
observations.  Section~\ref{sec.discussion} discusses various aspects
of this work, and Section~\ref{sec.conclusions} summarizes the paper.

\section{METHODS} %2
\label{sec.methods}
\subsection{Overview} %2.1

For reasons that will soon be apparent, a full description of our
analysis methods will be rather complex, so we begin by describing its
most essential aspects.  Let us first summarize the model in
\citetalias{fardal07}.  This starts with a satellite galaxy of a
certain mass and density profile, moving in a potential specified by
parameters for M31's baryonic and dark components, on a certain orbit
which happens to be highly radial.  After its first pass through
pericenter, its stars disperse onto orbits with a wide range of
energies, which spread and form structures resembling the GSS, NE
Shelf, and W Shelf features in M31.  (See Figure~\ref{fig.map}, and
compare the simulations in Figure~\ref{fig.morphology} which have
similar sky patterns to that in \citetalias{fardal07}.)  This single
model is qualitatively quite successful, as it simultaneously explains
many separate features in detail and has made several successful
predictions (see \citealp{gilbert09,fardal12}).  At the same time,
there are quantitative aspects that could be improved, particularly
the velocity of the GSS.

The \citetalias{fardal07} model is simple enough to be described just
by the parameters listed in Table~2 of that paper.  Some of these are
not of any fundamental interest, like the exact starting point and
speed of the satellite.  Others are more significant, such as those
describing the gravitational potential of M31.  We can think of the
\citetalias{fardal07} model as a single point in a low-dimensional
parameter space.  We would like to know, is there a better model
somewhere in the space?  And where do the allowable models live; more
precisely, what is the full probability distribution in this parameter
space?

Bayesian statistics gives us a means of answering these questions,
at the cost of supplying four separate ingredients.  
The first ingredient, which we already have,
is a model parameter space containing parameter vectors $\mathbf{M}$.
The next is a prior probability distribution $P(\mathbf{M})$
on this parameter space, which takes into account
information beyond that incorporated in the particular measurement we
are making.  For example, M31's gravitational potential is not known
precisely, but we can try to quantify its probability distribution
based on observables that have nothing to do with the GSS.  The next
ingredient is the likelihood function 
$\mathcal{L} = P(\mathbf{D}|\mathbf{M})$, 
the probability distribution of the GSS-related data given the model.
In our case this needs to be calculated by performing a
simulation with the given parameters, then comparing this simulation
to data in a quantitative way.  Using Bayes's theorem, we can write
the final (or posterior) probability distribution 
$P(\mathbf{M}|\mathbf{D})$ of the parameters as
the product of the prior and likelihood functions:
\begin{equation}
P(\mathbf{M}|\mathbf{D}) = \frac
{P(\mathbf{D}|\mathbf{M}) P(\mathbf{M})}
{P(\mathbf{D})}
\end{equation}
(In this paper we will generally regard ${P(\mathbf{D})}$ as an 
uninteresting normalization constant.)
The final ingredient in Bayesian statistics is an effective means of
sampling parameter vectors from the posterior probability
distribution, in order to describe the distribution accurately. 
As is often the case, we supply this using a
particular form of Markov Chain Monte Carlo (MCMC) sampling.
Having finally produced a sample of parameter vectors, we can
manipulate it in many ways to obtain physical information.  For
example, we can compute the marginal distribution for each of the
individual parameters, or for any quantity of interest that can
computed from them.

This paradigm of Bayesian sampling is probably familiar to most
readers.  There are some special features in our case, though, that
will make our procedure quite detailed and difficult.  One is
obvious: performing an $N$-body simulation at each likelihood evaluation
is rather slow.  Even if we sample using MCMC methods 
rather than a brute-force grid search, and split the calculation
into chains run in parallel, we still expect the sampling
will take hundreds or thousands of steps at a minimum.  So we will
need to perform the sampling efficiently to have any hope of success.
We also aim to keep the model space as simple as possible, since sampling 
the distribution becomes more difficult when more parameters are added.

Other difficulties arise from the particular problem studied here.  For
example, we need to combine a set of different types of observations to
effectively constrain the model, and for each type we explain in detail 
how we quantify its contribution to the likelihood function.  In addition, 
we must be careful about which observations {\em not} to include, either
because we believe they are contaminated by other components beyond
hope of recovery, or because we expect they cannot be reproduced
effectively by our simple models.  Complicating things further, we
intend to use several parameter spaces of different dimension,
in order to understand the effects of various parameters and the
robustness of our results.  

Furthermore, there is one feature of our sampling problem that seems
unusual, at least in an astronomical context.  
Our likelihood function is based on a particle simulation,
and therefore is {\em intrinsically} stochastic, 
significantly so in our case.\footnote{Even though our 
approach involves approximate Bayesian computation,
it should not be confused with the so-called 
Approximate Bayesian Computation method \citep[e.g.,][]{beaumont10}.
This latter approach dispenses with the mathematically formulated
likelihood altogether, and instead samples from parameter space
while requiring that a full simulation of the
data values, including observational errors, lie close (in some
predefined sense) to the observed data values.}
There is a true value of the model likelihood that
would be obtained with a large number of simulation particles, 
but the likelihood value we obtain is randomly scattered around it.
If we repeat the simulation, using 
a different random seed to initialize the satellite particles,
we will get a different value for the likelihood function.  
We discuss the effect of this stochasticity
or ``likelihood noise'' in Section~\ref{sec.sampling.noise} 
and suggest several rules of thumb for dealing with it.  The one to
remember for now is that we should try to keep the number of
observational constraints as low as possible, consistent with the
requirement that we effectively constrain the posterior distribution.

In the rest of this section we fill in the details of our procedure.
(The reader not interested in details might be well-served to glance
at Figure~\ref{fig.map} and Tables~\ref{table.spaces}--\ref{table.imdata},
and then skip to Section~\ref{sec.results}.)  In Section~\ref{sec.collision}
we describe our collision model, including the initial
structure of the satellite, the orbital initial conditions, and our
methods for initializing and evolving the system in an automated way.
We also describe our quasi-analytic approximation for the
stream's central path, which will be used in several parts of the
likelihood function.   We specify in Section~\ref{sec.parameters}
several related model parameter spaces from
which we will draw samples, and the priors on the parameters.  
In Section~\ref{sec.likelihood}, we then
discuss the different types of observational data used, and how we
compare to these to form the various terms in the likelihood function.  
Section~\ref{sec.sampling} describes our procedures for sampling from the 
posterior probability distribution, including methods to help deal with
noise in the likelihood function.  Finally, we describe in 
Section~\ref{sec.samples} the actual MCMC runs generated using these 
procedures.

\subsection{Collision model}  %2.2
\label{sec.collision}

\begin{figure*}
%\leavevmode \epsfysize=8cm \epsfbox{figures/rawmap.eps} 
%            \epsfysize=8cm \epsfbox{figures/regmap.eps}
\leavevmode \epsfysize=11.8cm \epsfbox{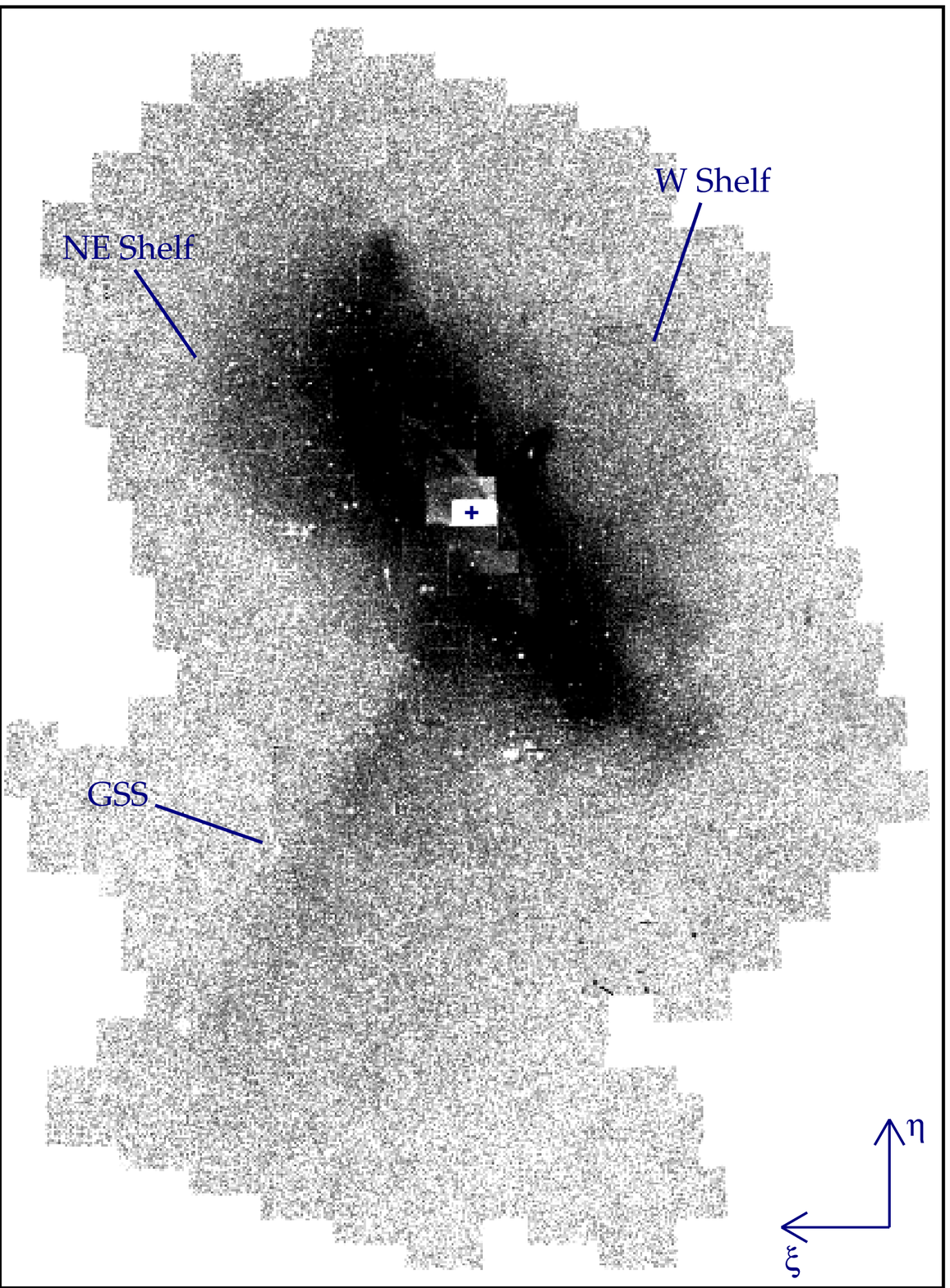} 
            \epsfysize=11.8cm \epsfbox{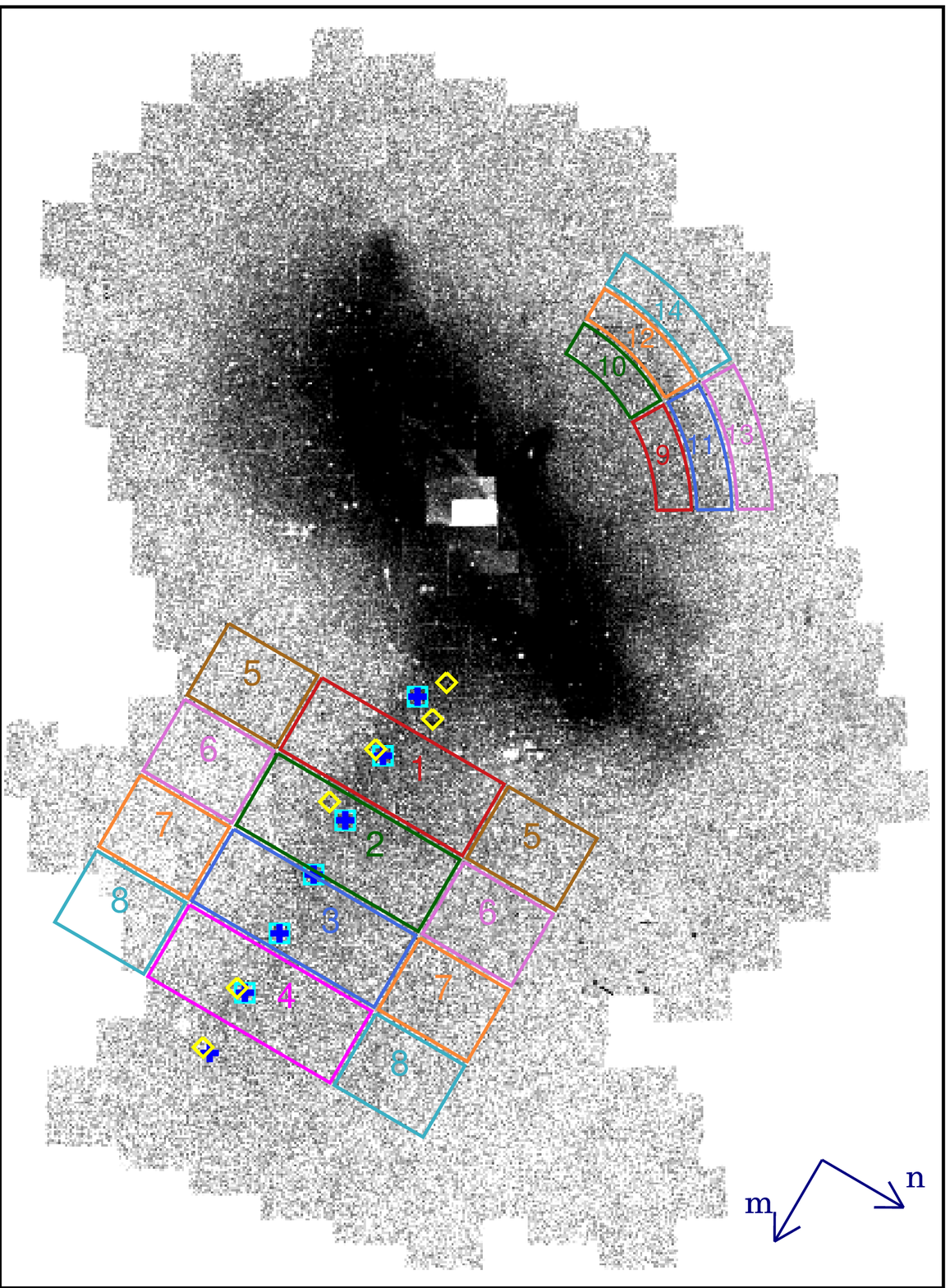}
\caption{
\label{fig.map}
Left panel: map of the RGB count density around M31 from the INT survey.
Pixel intensities show on a logarithmic scale the number of bright red
RGB stars within the color-magnitude cut described in the text.  
The main morphological features used to
constrain the GSS model are indicated, and M31's center marked with a cross. 
Right panel: same map, but showing regions used to constrain
the likelihood function.
Outlines with numbers indicate the regions used
to compare the observed count density
to mass surface densities in the simulations.
The ``outrigger'' regions in the GSS are split to either
side of the ``central'' regions.
Squares show the positions used for the central path of the stream,
which are barely offset from the field centers of \citet{alan03}.
Central positions of fields used for other stream measures are
shown by crosses for distances, and diamonds for stream velocities.
The maps use a tangent-plane projection centered on M31
with pixels 1 arcmin on a side.
They cover a range $3.5\deg$ to $-3.5\deg$ in $\xi$, and
$-5.75\deg$ to $3.75\deg$ in $\eta$, 
i.e.\ approximately $95 \kpc \times 130 \kpc$.
Axes at lower right in each panel show the orientations of the 
sky-aligned $\xi$-$\eta$ system and the GSS-aligned $m$-$n$ system
used in the paper.
}
\end{figure*}
 
Each of our parameter evaluations starts by 
setting up a self-gravitating \nbody\ model of a satellite in a
fixed potential model mimicking that of M31, infalling from near
apocenter.  We then follow its evolution with an \nbody\ code for a
time prescribed by the orbital parameters.  Our basic scenario (as in
\citetalias{fardal07}) specifies that as one follows the stream material
forward in its path, it forms the GSS region, NE Shelf, and W Shelf
(see Figure~\ref{fig.map}) from three successive orbital wraps.
%We follow the basic techniques of \citet{fardal06} and \citet{fardal07}.  

Here we use a spherical Plummer model for the satellite.  This is an
extremely simple model, which nevertheless provides a reasonable fit
to many real low-mass galaxies \citep{alan06}.
%Note: higher-mass reference would be nice but couldn't find one.
%Most used for dSph, but doesn't look so bad for many of the dIrr 
%with shallow inner profiles, at least within 2*plummer radius.
%see fingerhut et al 2010, hunter and elmegreen 2006 for some profile examples.
%the ``sech profile'' favored by fingerhut looks a lot like plummer except in
%the outer power-law tail.
%Of course Plummer more favored by theorists than observers...
In \citet{fardal08} we studied the effect of a flattened rotating
progenitor, finding the final overall morphology was generally
similar.  However, there are some subtle differences from a spherical
model, chief among them the skewed transverse distribution of stars in
the GSS.  Later we will design the likelihood function to be
insensitive to this skew.  For the purposes of this paper, the
simplicity of the spherical model outweighs anything to be gained by
using a more complex rotating model.

Real galaxies should have at at least two differently distributed
components, stars and dark matter, and perhaps a gas component as
well.  However, the stars in the GSS are almost entirely $>4$~Gyr 
old with a typical age of $\tsim 8$~Gyr,
as found by modeling of HST/ACS data on the GSS down to the main
sequence in \citet{brown06b}. This suggests there has been little gas 
in the satellite for some time.  This lack of gas might be related to 
the length of time the satellite has spent in M31's halo.  A typical
apocenter for the GSS progenitor of $\tsim 50 \kpc$, as found in 
\citetalias{fardal07}, is also inconsistent with a very recent
accretion from outside the virial radius.  Therefore we expect that
much of the progenitor's dark matter will have been stripped off in
orbits prior to its most recent collision, consistent with trends found
in cosmological simulations \citep{libeskind11}.  We thus find it both
reasonable and convenient to assume a single $M/L$ ratio throughout
the satellite.  To us it seems equally plausible that the mass should
be dominated by stars or dominated by dark matter, and we consider
two models, one where we fix $M/L$ and one where we leave it free to vary.  
\citet{mori08}
have considered a GSS progenitor with a massive halo, finding that a
total satellite mass of $\ltrsim 5 \times 10^9 \msun$ is required to avoid
perturbing M31's disk too much.  We will find later that much of the mass
allowed by this limit will be taken up by stars, leaving little room 
for a dark halo.

We create the initial satellite particle distribution with the library
ZENO by Josh Barnes.  This solves the Abel integral to compute the
exact velocity distribution for the specified density model, enabling
rapid production of accurate equilibrium initial conditions for any
initial profile, though in this paper we restrict ourselves to the
Plummer profile which has an analytic solution. We use 65,536
particles in each satellite, as a compromise between the competing
demands of computation time and simulation noise (see the discussion
in Section~\ref{sec.sampling}).  Tests with an isolated satellite show
minimal evolution over the typical duration of the run.  Each
simulation uses a different random number seed, and thus a different
distribution of particles even for identical parameters.

We omit dynamical friction from the calculation.  In earlier test
simulations with live halos and disks and analytic approximations, we
found that the main effect of dynamical friction in reasonable models
is to induce a loss of energy at the first pericentric passage, when
the satellite is still intact.  This means the orbit should actually
be started at a slightly higher energy, but otherwise the structure is
little affected.  It would be impractical to conduct a large number of
simulations with a full $N$-body model for the primary.  Analytic
approaches can provide at least an estimate of the drag on the
satellite, but it is difficult to accurately calculate the force field
that should be applied to the entire tidal structure.  In addition, it
simplifies the analysis to be able to speak of fixed orbital
quantities such as the apocenter.  Therefore we postpone consideration
of dynamical friction to future work.

In some tidal stream problems it is convenient to simplify the problem
even further using a reduced \nbody\ model, i.e. fix the satellite
potential and follow the evolution of massless test particles.
However, in our case the self-gravity is essential.  Clearly the
satellite is largely ripped up by its encounter, so the time variation in
its potential has a major effect on the final debris.  Comparison of
full and reduced \nbody\ calculations show a large difference in the
stream's surface density in particular. Since the stellar surface
density will be one of the primary observables, it is essential to use
a live satellite despite the extra computional burden.

We set up a fixed M31 potential consisting of a bulge, disk, and halo,
where the parameters controlling this potential are described in more 
detail later.  We follow the satellite's evolution with PKDGRAV, a 
versatile multi-stepping tree code written by Joachim Stadel and Tom Quinn
\citep{stadel01thesis}. We have customized this code to accept
parameters for our fixed potential components.  We choose a spline
softening length equal to 0.3 times the Plummer scale radius.  We also
use fourth-order multipoles in the node potential expansions, set the
opening angle to $\theta = 0.8$, and use the acceleration-based
timestepping criterion with $\eta_{acc} = 0.2$.

\subsubsection{Orbital approximation} 
\label{sec.orbital}
As mentioned before, the orbit of the GSS progenitor is a very poor
approximation to the resulting stream \citep{fardal06}, but it is
possible to find an approximation of the central path of the stream in
phase space.  The most useful approximation in the current case, where
the orbit is highly radial and the progenitor disrupts almost
completely at pericenter, remains that of \citet{fardal06}, which
relies on scaling the orbit to obtain a track of the stream.  While we
now wish to use simulations to compare to observations, this
``stream-orbit'' approximation, so called because it directly relates
the progenitor orbit to the path of the stream, remains useful in
several respects.  First, we can construct a likelihood function based
entirely on a single orbital calculation and then sample the resulting
probability distribution with standard Bayesian techniques.  The
resulting ``orbital'' sample will not reflect all the observational
constraints, but will be useful as a starting point for the
simulation-based sampling chains.  Second, since noise in the
likelihood function will prove to be an important factor, we will
actually use the orbital model in place of the noisy simulations to
estimate several of the stream quantities when computing the full,
simulation-based likelihood.

The stream locus in phase space is obtained by first calculating
the progenitor's trajectory, assuming it behaves as a test particle
in the potential of M31, then analytically distorting it.
The orbit is described
by position $\mathbf{r_p}(t)$ and velocity $\mathbf{v_p}(t)$,
We assume that the potential of M31 is roughly
spherical and close to a power-law $\Phi \propto r^k$.  In
\citet{fardal06} we assumed a constant exponent $k=-0.4$,
significantly steeper than an isothermal halo.  Here, we instead
calculate $k$ from the ratio of the gravitational force at 15 and
45~kpc.  Given our adjustable potential, this change leads to better
agreement with the simulated velocities.  
For a range of values of $M_{200}$
from $7 \times 10^{11}$ to $3 \times 10^{12} \msun$, this value ranges
from $-0.4$ to $+0.1$.

We assume that the stars in the stream were all liberated at the 
first pericentric passage, an assumption valid for a highly radial
orbit, and set $t=t_d$ at this point.  Then we assume that the different
stars in the stream have orbits that are geometrically similar but
scaled in orbital period by a factor $\tau$.  Constructing an array of
$\tau$ values over some large range, we can then construct a
parametric solution of the stream locus at time $t$
by setting the position to
$\mathbf{r} = \tau^{2/(2-k)} \mathbf{r_p}[(t-t_d)/\tau]$,
and the velocity to
$\mathbf{v} = \tau^{k/(2-k)} \mathbf{v_p}[(t-t_d)/\tau]$
(see \citealp{fardal06}).  

When we later obtain actual simulations drawn from various parameter
samples, we find good overall agreement between the simulations and
the stream-orbit model.  Later we discuss slight correction terms we
include to improve the model for quantities used in the likelihood
function.

\subsection{Model parameters and priors} %2.3
\label{sec.parameters}

Our goal in constructing the collision model has been to include
parameters that have a large effect on the results, while keeping the
dimensionality low.  To test the significance of different parameters,
we define five different parameter spaces in this paper,
called the ``stellar'', ``DM'', ``orbital'', ``reduced'', and
``density'' spaces.  
The parameters used in these model spaces are summarized
in Table~\ref{table.spaces}.
We begin by discussing the two largest dimension
spaces, ``stellar'' and ``DM'', which are the main spaces used for
science results.
As each parameter is introduced, we discuss its prior probability
distribution; these priors are summarized in Table~\ref{table.paramlimits}.

We state our orbital parameters in a Cartesian coordinate system
centered on M31 and aligned with the sky, although the actual
calculation uses a rotated system aligned with M31's disk.  The sky
system is chosen so that $X$ increases with right ascension to the E,
$Y$ increases with declination to the N, and $Z$ increases with
distance from Earth.  It is simple to translate these into the tangent
plane coordinates $\xi$ and $\eta$ (aligned with $X$ and $Y$
respectively), distance $d$, and velocity $v_r$
\citep[see][]{fardal06}.  
%We use M31-centered distance $d_{M31} = d-780 \kpc$ and velocity
%$v_{M31} = v_r + 300 \kms$ in later plots.

The orbital initial conditions require six free parameters.  As in
previous papers, we choose one of these to be the orbital phase of the
progenitor at the present day.  We express this phase $F_p$ in terms
of the time since the initial disruptive pericentric passage (at
$t_d$) in units of the progenitor's radial period $t_r$, so that 
$F_p = (t-t_d) / t_r$.  Models with similar $F_p$ tend to resemble each 
other except for differences in orientation and scale.
In our scenario, $F_p=0$ represents the initial disruption point, 
$F_p=1$ implies the progenitor has 
just completed the GSS part of the orbit, $F_p=2$ means it
has just moved out to the NE Shelf and back to pericenter, and $F_p=3$
means it has just completed the W Shelf part of its orbit and
returned to pericenter.
Since observations suggest that the stream density rises towards
the NE Shelf and falls off again in the W Shelf, 
we assume limits of 0.8 to 2.0.  We take the prior on $F_p$ to be uniform  
within these limits since the disruption time is arbitrary.  
Note the limits for all parameters are given in Table~\ref{table.paramlimits}.

With the addition of $F_p$, we then have one spare phase-space 
parameter, so we calculate the
orbit from a fixed plane $Y_0 = -10 \kpc$, which will lie within the
GSS orbital wrap.  We then calculate the orbit forward and backward
from this reference point.  The other parameters $X_0$, $Z_0$,
$V_{X0}$, $V_{Y0}$, and $V_{Z0}$ are all set to have uniform priors,
inside a box placed loosely around values from earlier successful
models rather than from any physical considerations.  We set 
$V_{Y0} > 0$ at this reference point so that it corresponds to an
upward-moving satellite at a time later than $t_d$, moving like the
stars in the GSS.  The satellite itself is started at a much earlier
point, just after apocenter on the previous orbital wrap at a time
before $t_d$.

In all except our ``density'' parameter space,
we scale the Plummer scale length of the initial
satellite according to $a = 0.8 \kpc (M_{sat}/10^9 \msun)^{1/3}$, so that
the central density is fixed at 
$\rho_{sat} = 4.6 \times 10^8 \msun \kpc^{-3}$ and the central surface 
density at $5.0 \times 10^8 \msun \kpc^{-2} (M_{sat}/10^9 \msun)^{1/3}$.
This agrees well with relations for local galaxies \citep{dekel03,woo08}. 
The mass $M_{sat}$ of the satellite is taken to have a uniform prior in 
its logarithm, from $10^{8.5}$ to $10^{10} \msun$ in our ``stellar'' model, 
and over a larger range from $10^{8.5}$ to $10^{11} \msun$ in our ``DM'' model.

We use the bulge-disk-halo potential model of \citet{geehan06}.  
The model contains a spherical Hernquist bulge, exponential or
Miyamoto disk, and spherical Navarro, Frenk, and White (NFW) halo.
As used here, this is a single-parameter family of potentials 
(the vertical line in Figure~6 of \citealp{geehan06}), with
the single degree of freedom mostly controlling the outer halo potential.
Two parameters in this model (bulge radius, disk scale length) are
well constrained, and we have fixed them at their best-fit values:
$r_b = 0.61 \kpc$, $r_d = 5.40 \kpc$.
The classic disk-halo degeneracy is a major degree of freedom in the
model; to resolve this we assume a central disk surface density of
$4.0 \times 10^8 \msun \kpc^{-2}$, corresponding to a disk mass
of $M_d = 7.34 \times 10^{10} \msun$ and $(M/L_R)_d = 3.3$.  
This model also contains a great deal of flexibility in the
halo potential because it is constrained only by relatively few
statistical tracers, namely PNe, globular clusters, and satellite
galaxies combined according to the Bayesian modeling of
\citet{evans00a} and \citet{evans00b}.  Within the code we
parameterize the halo's mass and scale
radius $r_h$ by $f_h = r_h / 7.9 \kpc$, since $7.9 \kpc$ was the scale
radius with the highest likelihood at our chosen surface density
\citep{geehan06}.  (In \citealp{fardal06} we used an earlier version 
of the \citealp{geehan06} model based on slightly different data,
leading to a maximum likelihood value for $r_h$ of $9.0 \kpc$.)
To within a few percent, the best-fitting bulge mass and halo density
parameter along this path are then given by
\begin{eqnarray*}
M_b & = & (3.24 + 0.238 u - 0.103 u^2 + 0.0158 u^3) \times 10^{10} \msun \; ,\\
\delta_c &= & \exp(12.66 - 1.956 u + 0.143 u^2) \; ,
\end{eqnarray*}
with $u \equiv \ln f_h$.
The halo density parameter in turn sets the dark matter NFW density profile
according to $\rho_h(r) = \delta_c \rho_c x^{-1} (x+1)^{-2}$, 
where $x = r / r_h$ 
and $\rho_c = 140 \msun \kpc^{-3}$ is the present-day critical density
for our assumed Hubble parameter $H_0 = 71 \kms \Mpc^{-1}$.
%  $\rho_0 = \delta_c \rho_c = 1.30 \times 10^4 \msun \kpc^{-3}$

In this paper we report the halo mass in terms of the ``virial mass''
estimator $M_{200}$, the mass inside a sphere containing an average
density 200 times the closure density of the universe.  We choose
this threshold in preference to $M_{100}$, also commonly adopted,
because of its closer relationship to the length scales actually
probed by the GSS.   
In our model the relationship of this to $f_h$ is described well by
\begin{eqnarray*}
\log_{10}(M_{200}/\msun) & = & 11.572 + 0.324 f_h - 0.0481 f_h^2 \\
                   &   & + 0.00438 f_h^3 - 0.000159 f_h^4 \; .
\end{eqnarray*}

Since speed is paramount in our case, we have slightly altered the
model to use a Miyamoto-Nagai instead of an exponential disk, with
$M_{miya} = 1.1 M_{exp}$ and $R_{miya} = 1.1 R_{exp}$, and disk scale
height $b_{miya} = 0.4 \kpc$.  This slightly changes the total mass
by an amount well within the uncertainty, and otherwise seems to give  
good agreement with the exponential disk forces at most positions.

We incorporate prior knowledge of the M31 halo potential into our
calculation.  The simple observation that M31 has baryonic mass $\tsim
10^{11} \msun$ already introduces a fair amount of prior knowledge.
Taking into account both the cosmic distribution of galactic halos ($dN/d\ln
M_{200} \propto M_{200}^{-0.8}$ in this mass range) and fairly generous
limits on the halo mass required to assemble this amount of baryons
into a galaxy, we adopt a Gaussian prior for $\log_{10} (M_{200}/\msun)$ with
mean 12.2 and dispersion 0.4 dex, which gives a FWHM of almost 1 dex.
We then multiply this by the likelihood value from the \citet{geehan06}
model to form our final prior on $M_{200}$.  
The Geehan et al.\ likelihood factor is quite strongly discouraging
of large halo values, dropping to 0.1 of its maximum value already at
$\log_{10} (M_{200}/\msun) = 12.1$ and declining exponentially with scale length
0.08 dex at higher values.  However, our previous experience suggests
that a good fit to the GSS debris is likely to be found for 
$f_h > 1$, so we restrict the run to the range $f_h = 0.8$--8, 
or $M_{200} = 0.63$--$4.7 \times 10^{12} \msun$.  
Purely for technical reasons, within our code we have specified the
prior as a flat distribution in $f_h$ and moved the remaining factors
to the likelihood.  Since the code ultimately samples from the
posterior (product of prior and likelihood), this modification has
no consequence for our results.  The prior distribution on $M_{200}$ can be
seen in the inset of Figure~\ref{fig.contours_st}; it is strongly 
biased to lower values, whereas the posterior obtained from our MCMC results
will have a very different mean.

\begin{table}
\caption{Parameter spaces}
\label{table.spaces}
\begin{tabular}{@{}ll}
\hline
Model space & Parameters \\
\hline
``stellar'' &
$X_0$, $Z_0$, $V_{X0}$, $V_{Y0}$, $V_{Z0}$, 
$F_p$, $M_{200}$, $M_{sat}$\\
``DM'' &
$X_0$, $Z_0$, $V_{X0}$, $V_{Y0}$, $V_{Z0}$, 
$F_p$, $M_{200}$, $M_{sat}$, $M_{lum}$\\
``orbital'' &
$X_0$, $Z_0$, $V_{X0}$, $V_{Y0}$, $V_{Z0}$, 
$F_p$, $M_{200}$ \\
``reduced'' &
$F_p$, $M_{200}$, $M_{sat}$\\
``density'' &
$F_p$, $M_{200}$, $M_{sat}$, $\rho_{sat}$\\
\hline
\end{tabular}

\medskip
$X_0$, $Y_0$, $Z_0$, $V_{X0}$, $V_{Y0}$, and $V_{Z0}$ denote
the phase space coordinates of the reference point in the 
progenitor's orbit.  
The orbital coordinate $Y_0$ is held fixed at $-10 \kpc$,
so it is not a free parameter.
$F_p$ is the current orbital phase of progenitor on it
by its initial orbit, expressed as the time since its
disruptive pericentric passage in units of the radial period.
$M_{200}$ is the halo mass of M31, measured within a radius where
the mean enclosed density equals 200 times the critical density
of the universe.  
$M_{sat}$ is the initial mass of the progenitor satellite,
and $M_{lum}$ its initial luminous mass in the model where these differ.
Finally, $\rho_{sat}$ is the central mass density of the progenitor satellite.
%\end{minipage}
\end{table}

\begin{table}
\caption{Parameter priors}
\label{table.paramlimits}
%\begin{tabular}{@{}cccc}
\begin{tabular}{@{}c@{}c@{\hspace{1em}}c@{}c}
\hline
Parameter&Minimum&Maximum&Distribution \\
\hline
$X_0$ ($\kpc$)    &  $-15$  &  20    &  uniform\\
$Z_0$ ($\kpc$)    &  1  &  50    &  uniform\\
$V_{X0}$ ($\kms$) &  $-200$  &  $-10$    &  uniform\\
$V_{Y0}$  ($\kms$) &  10  &  380    &  uniform	\\
$V_{Z0}$ ($\kms$) &  $-380$  &  $-100$    &  uniform\\
$F_p$            &  0.8  &  2.0    &  uniform	\\
$\log_{10} (M_{200} / M_{\sun})$  &  11.80  & 12.67  &  peaked, see text\\
$\log_{10} (M_{sat} / M_{\sun})$&  8.5  &  10.5    &uniform (``stellar'')\\
$\log_{10} (M_{sat} / M_{\sun})$&  8.5  &  11.0    &  uniform (``DM'')\\
$\log_{10} (M_{lum} / M_{sat})$&  $-1.0$  &  0.5    &  uniform\\
\hline
\end{tabular}
%\end{minipage}
\end{table}

This completes the specification of our ``stellar'' model space, which
assumes the satellite's mass is entirely stellar in nature.  For our
``DM'' model, we draw a distinction between the luminous (stellar)
mass $M_{lum}$ and the total mass $M_{sat}$, which we assume to be
blended with (though not necessarily dominated by) dark matter.  
We parameterize this by the ratio
$M_{lum}/M_{sat}$, for which we take a flat logarithmic prior, and
we allow this to vary over a large range (even to values above 1,
to see if the observations suggest an error in our mass calibration 
derived later).  

As mentioned earlier, we define an ``orbital model'' which is based
only on the stream-orbit calculations of Section~\ref{sec.orbital}.
Since the satellite mass or debris distribution is not considered
in this model, the ``orbital'' space is equivalent to the ``stellar''
space with satellite mass omitted.

We also consider a ``reduced'' parameter space, where we eliminate all
orbital parameters except for the orbital phase $F_p$.  Here we use
the orbital model to compute the maximum-likelihood value of the other
orbital parameters $X_0$, $Z_0$, $V_{X0}$, $V_{Y0}$, and $V_{Z0}$,
over a grid of $F_p$ and $M_{200}$.  This leaves a three-dimensional
space of $F_p$, $M_{200}$, and $M_{sat}$, which can be sampled more
easily than our standard model.  We used this space for testing
purposes while working towards the larger-dimensional results.

Finally, we defined a ``density'' space, an augmented version of the
``reduced'' space in which the central density of the progenitor's Plummer 
model $\rho_{sat} = (3 M_{sat}) / (4 \pi a^3)$
was left free to vary.  We imposed a Gaussian prior on 
$\log_{10} \rho_{sat}$ with mean $-0.3$ dex, consistent with the 
fixed value in our main samples, and dispersion $0.5$ dex.  
We used this space later to test the influence of adding a freely 
varying satellite size scale, without the computational burden of the 
full parameter space.

\subsection{Likelihood function}  %2.4
\label{sec.likelihood}

Now that we have explained how we go from a point in parameter space
to an \nbody\ debris structure, we need to specify the likelihood function,
the probability of drawing the data from this debris structure given
the \nbody\ model regarded as a function of the model parameters.  
Because M31 is
a highly inhomogeneous environment with many overlapping structures,
we try to use only the parts of the data that are least contaminated
by unrelated material.  As mentioned earlier, each term in the 
likelihood function that is calculated from the simulation adds noise 
or stochasticity to the likelihood.  
Thus it is helpful to keep the number of terms 
using the simulation on the low side, while still aiming at a
combination of data that will constrain the model.  

Rather than the true likelihood $\mathcal{L}(D|M)$ (where 
$D$ means data and $M$ means model),  
it is usually easier to work with its logarithm $L(D|M) = \ln(\mathcal{L})$.  
This log-likelihood is comprised of several terms corresponding
to different types of GSS observations.
Some of these terms were previously used in
the orbital fitting of \citet{fardal06} and \citetalias{fardal07}, though we have
slightly updated the data values:
\begin{itemize}
\item[-] $L_{pos}$, which constrains the central position of the GSS;
\item[-] $L_{d}$, which specifies the distance of the GSS;
\item[-] $L_{v}$, which specifies the velocity of the GSS;
\item[-] $L_{lobe}$, which loosely describes the position of the NE shelf.
\end{itemize}
To this we have grafted on 
\begin{itemize}
\item[-] $L_{\im}$, which describes the surface density pattern of stream
material in the GSS and W shelf regions, as based on the image in 
Figure~\ref{fig.map}. 
\end{itemize}
All the terms in our likelihood happen to be of the $\chi^2$ form, with 
$\mathcal{L}_i \propto \exp[-\Sigma_i (D_i-M_i)^2/(2 \sigma_i^2)]$, so that
$L = -\chi^2 / 2$ with $\chi^2 \equiv \Sigma_i (D_i-M_i)^2/\sigma_i^2$ for 
independent data points, or 
$\chi^2 \equiv (\mathbf{D}-\mathbf{M}) \mathbf{C}^{-1} (\mathbf{D}-\mathbf{M})$ 
for covariant data points where $\mathbf{C}$ is the covariance matrix.  

\begin{table}
\caption{Observational data}
\label{table.obsdata}
%\begin{tabular}{@{\hspace{1em}}l@{\hspace{1em}}c@{\hspace{1em}}c@{\hspace{1em}}c@{\hspace{1em}}c@{\hspace{1em}}l@{\hspace{1em}}}
\begin{tabular}{@{\hspace{.6em}}l@{\hspace{.8em}}c@{\hspace{.8em}}c@{\hspace{.8em}}c@{\hspace{.8em}}c@{\hspace{.8em}}l@{\hspace{.6em}}}
\hline
Field & $\xi$ & $\eta$ & Mean  &  Error & Reference \\
\hline
\multicolumn{6}{c}{Stream transverse coordinate $n$ (degrees)}\\
F2 & 1.68 & $-3.56$ &0.34  &  0.20  & \citet{alan03} \\
F3 & 1.43 & $-3.12$ &0.34  &  0.20  & and see text \\
F4 & 1.17 & $-2.68$ &0.34  &  0.20  & \\
F5 & 0.94 & $-2.28$ &0.34  &  0.20  & \\
F6 & 0.66 & $-1.80$ &0.34  &  0.20  & \\
F7 & 0.40 & $-1.37$ &0.34  &  0.20  & \\
\hline                 
\multicolumn{6}{c}{Stream distances (kpc), large to small radius}\\
F1 & 1.94  &   $-4.01$  &  886  &  20  &  \citet{alan03}\\
F2 & 1.68  &   $-3.56$  &  877  &  20  &   \\
F3 & 1.43  &   $-3.12$  &  860  &  20  &   \\
F4 & 1.17  &   $-2.68$  &  855  &  20  &   \\
F5 & 0.94  &   $-2.28$  &  840  &  20  &   \\
F6 & 0.66  &   $-1.80$  &  836  &  20  &   \\
F7 & 0.40  &   $-1.37$  &  829  &  20  &   \\
\hline
\multicolumn{6}{c}{Stream velocities ($\mbox{km}\,\mbox{s}^{-1}$), large to small radius} \\
s1  & 1.99  & $-3.97$  &  $-18$    &  25 &  \citet{ibata04} \\
s2  & 1.74  & $-3.52$  &  $-45$    &  25 &  \citet{ibata04} \\
a3  & 1.06  & $-2.14$  &  $-141$ &  8  &  \citet{gilbert09} \\
s6  & 0.71  &  $-1.76$ &  $-181$   &  25 &  \citet{ibata04} \\
H13s & 0.29  & $-1.53$ & $-190$ &  8  &  \citet{gilbert09} \\
f207 & 0.19 & $-1.26$ & $-224$ &  10 &  \citet{gilbert09} \\
\hline
\end{tabular}
%\end{minipage}
\end{table}

\begin{table}
\caption{INT image data}
\label{table.imdata}
\begin{tabular}{@{}ccccc}
\hline
Region & Raw &  Background  & Background & Area \\
           & counts     &  counts  & error & \\
           & \multicolumn{3}{c}{\dotfill (counts arcmin$^{-2}$) \dotfill} &  (arcmin$^2$)\\
& & & & \\
\hline
  1  &  3.49  &  1.54  &  0.13 &  3744\\
  2  &  2.58  &  1.19  &  0.09 &  3744\\
  3  &  2.12  &  0.98  &  0.03 &  3744\\
  4  &  1.76  &  0.91  &  0.05 &  3744\\
  5  &  1.72  &  1.44  &  0.10 &  3744\\
  6  &  1.42  &  1.17  &  0.07 &  3744\\
  7  &  1.31  &  0.99  &  0.05 &  3744\\
  8  &  1.30  &  0.94  &  0.06 &  3744\\
  9  &  3.29  &  2.18  &  0.27 &   848\\
 10  &  3.20  &  2.09  &  0.25 &   848\\
 11  &  2.47  &  1.69  &  0.15 &  1018\\
 12  &  2.49  &  1.71  &  0.16 &  1018\\
 13  &  1.35  &  1.54  &  0.14 &  1188\\
 14  &  1.59  &  1.59  &  0.13 &  1188\\
\hline
\end{tabular}
%\end{minipage}

\medskip
Regions are indicated in Figure~\ref{fig.map}.  
\end{table}

\subsubsection{Surface density}
To constrain the surface density of GSS debris, we use the Isaac
Newton Telescope (INT) {\it Wide-Field Camera} survey of M31's halo in
Johnson $V$ and Gunn $i$ filters \citep{ibata01,ferguson02,irwin05}.
This survey is deep enough to reach M$_V \approx 0$ on the RGB for
stars in M31.  Details of the photometric pipeline and morphological
classification are described in \cite{irwin01}.  Color-magnitude
cuts in the survey catalog are effective in picking out M31 RGB stars
and revealing smooth and structured components of M31's halo, although
foreground Milky Way halo and disk stars and background galaxies are
both still major contaminants.  The deeper and broader PAndAS survey
\citep{alan09} (which builds on the quadrant surveyed earlier by
\citealt{ibata07}) recently finished acquiring data, but final analysis
of the dataset is still ongoing.  Fortunately, the INT survey has
enough source counts to strongly constrain our models.

The GSS is known to be relatively metal-rich, and therefore most
easily visible in red RGB stars, as seen in the map of
\citet{irwin05}.  However, the metallicity distribution within the
stream is fairly broad \citep{ibata07}, so that it is also
apparent in the corresponding map of lower-metallicity blue RGB stars
in \citet{ferguson02}.  Therefore we use a color cut that is slightly broader 
than in Irwin et~al.'s ``red'' RGB sample.  
We first deredden the source magnitudes for Galactic extinction
to $V_0, i_0$ according to \citet{schlegel98}.  We then use cuts of
$21 < i_0 < 22$, $ V_0-i_0 > 1.30 + 0.35 (22 - i_0)$, 
representing a compromise
between maximizing source counts and excluding unrelated
lower-metallicity structures.  We include only sources 
with magnitude uncertainties less than 0.25, and classified as
``stellar'' or ``probably stellar'' by the survey pipeline.  
Published INT maps show square artifacts from field-to-field variations,
presumably due to the effect of seeing variations on source detection
and star/galaxy separation.  A significant fraction of sources
satisfying the $i$ magnitude cut have no detection in $V$, and these
show the same square artifacts. We include these sources in our count map
weighting each by 50\%, as we found that produced smoother maps
compared to either including or excluding them entirely.  
After mapping sources onto a tangent-plane projection and 
binning into 1 arcmin square pixels, we obtain the star-count
map shown in the left panel of Figure~\ref{fig.map}.

In our scenario for the GSS's formation, we expect that moving the
core forward in its orbit (i.e.\ increasing the progenitor phase
$F_p$) will cause the surface density in the GSS to decrease, while
the opposite is true for the W Shelf which lies ahead of the core (as
long as $0.8 < F_p < 2.0$).  In \citetalias{fardal07} we suggested
that the ratio of surface brightness in the GSS and W Shelf would make
a good constraint on $F_p$.  Here we put this idea into practice.  We
also use the star-count map to limit the width of the simulated GSS,
which relates to its mass and the impact parameter of the orbit.

We choose a set of bins on the sky to satisfy these goals, while
avoiding regions like the NE Shelf that are mixed with M31's disk
(see Figure~\ref{fig.map}, right panel).  
We set up a rectangle covering the GSS region, slice it in four
pieces lined up along the stream, then cut each slice into one
central and two ``outrigger'' bins so that the central bin
contains 50\% of the slice's area.  The range along the stream
is chosen to avoid both the outer regions where the GSS signal is very
faint, and the inner regions that for some parameter settings are covered 
by the NE Shelf orbital wrap.  We anticipate from our work on rotating
GSS progenitors \citep{fardal08} that the spherical, non-rotating
progenitors used here will have difficulty matching the skewness of
the transverse distribution.  Therefore we combine the counts of the
satellite regions on each side of each central region to form a single
bin.  To first order this should be unaffected by skewness of the
transverse distribution.

In the W Shelf we use three thin radial cuts, with the central one
containing the visible W Shelf edge, each split down the middle 
into two bins to 
loosely constrain the shelf's azimuthal position.  Therefore, in
addition to measuring the overall count level and the ratio of GSS to 
W Shelf counts, the overall combination of 14 bins also measures the
gradient of surface density in the GSS, the width of the GSS, and the
location of the W Shelf.  

We now need to obtain the data values in each bin.  We boxcar smooth
the star-count map with a window 20 arcmin on a side, and mask out
small gaps and outlier pixels by comparing the smoothed and unsmoothed
maps.  We compute the raw data counts per pixel $\mathbf{D}_{raw}$ by
taking the usable map pixels within each bin shown in
Figure~\ref{fig.map} and averaging their counts.

We next must estimate the mean, uncertainty, and covariance matrix
of the contribution from ``background'' sources unrelated to the 
GSS.\footnote{Our use of the term ``background'' is not meant to 
suggest anything about the spatial position of the contaminating sources, 
most of which indeed are foreground Milky Way stars.}
To do so, we first select regions in the extreme N and S far from 
M31 and its substructure, fit a linear model in $\eta$ (which varies
nearly parallel to galactic latitude $b$ in the survey region) to estimate
the Milky Way and background galaxy contribution, and subtract this off
to eliminate contaminants present across the entire map.  
This leaves only the contamination from M31's smooth and tidal components.

To estimate this latter source of contamination, we first define 
a region relatively free of tidal features as follows.  
We define ellipses aligned with M31's disk following the 
estimated halo axis ratio of 0.6 \citep{pritchet94}.
Marching outwards in elliptical annuli, we mask out pixels 
in the smoothed map that are $5 \sigma$ above the mean annulus 
value, then recalculate the mean and repeat the masking.  
After also masking any pixels within the test bins shown in
Figure~\ref{fig.map}, we obtain a background estimation region where
the GSS, NE and W shelves, and several other tidal features have been
masked out.

We assume the M31 ``background stars'' are roughly symmetric
about M31's center.  To estimate the mean background value 
in each bin and the covariance between bins, then, we therefore
spin the entire bin pattern around M31 and measure the mean and
covariance values averaged over spin angle.
Since we are assuming an average axial ratio of 0.6 for the background
component, we distort the bins to follow ellipses of this shape.
We use 30 evenly spaced orientations, and 
exclude the masked region from the measurements. 
For each bin and orientation we measure the counts per usable pixel.
Averaging over orientations yields our mean M31 background estimate
for each bin.  We also measure the dispersion in values from different
orientations, which becomes the background uncertainty $\sigma_{bg,i}$,
so that the background covariance matrix on the diagonal is 
$C_{bg,ii} = \sigma^2_{bg,i}$.
Estimating the covariance of background estimates in pairs of 
bins is more difficult.  
For each pair of bins we measured the cross-correlation, but most
pairs have only a few spin orientations where both bins are in the
background test region making this an unreliable estimate.
Combining the sample of pairs, we find the median correlation coefficient 
of neighboring bins (those sharing a common edge) is about 20\%.
We therefore set the covariance matrix coefficient for neighboring 
bin pairs to 20\%\ times the measured bin dispersions: 
$C_{bg,ij} = 0.2 \sigma_{bg,i} \sigma_{bg,j}$.
We zero out all covariance matrix elements for non-neighboring bins.
We compute the total covariance matrix $\mathbf{C}_{\im}$ by placing
the combined shot noise variance of the observation and simulation 
terms onto the diagonal, adding a diagonal error term 
corresponding to 10\% variation in the total counts
(to generously account for the influence of the artifacts discussed above), 
and adding these to the background covariance matrix $\mathbf{C}_{bg}$.  
Typically the background fluctuations and systematic error terms are 
the largest error source on the diagonal of $\mathbf{C}_{bg}$, with the 
simulation shot noise next and the observed shot noise the smallest.
 
We then add back the smooth Milky Way and galaxy term to obtain
the total background mean $\mathbf{D}_{bg}$.
Finally, we subtract this from the raw bin values to obtain our signal
estimate $\mathbf{D}_{\im} = \mathbf{D}_{raw} - \mathbf{D}_{bg}$.  For
the bins with obvious GSS contribution, this signal is 
$\gtrsim 5\sigma$ above the background level.  The raw count values
$D_{raw,i}$, background values $D_{bg,i}$, and background
uncertainties $\sigma_{bg,i}$ are listed in Table~\ref{table.imdata}.
We have investigated alterations to this procedure, including the
choice of spherical symmetry for the background calculation and
changes in the number of orientations or coverage maps, and generally
find quite similar results for the means and dispersions of the
background counts in the bins.  We thus have some confidence that our
background characterization is reasonable.

Next, we must calculate the effective number counts due to the simulation,
$M_{\im}$.
We calibrate star counts to mass using a theoretical stellar population 
based on observations of the GSS itself.
Our goal is $m_1$, the total stellar and stellar remnant mass 
associated with the detection
of one M31 star within our color-magnitude cut.
We first convert our INT cut to Johnson $V$, $I$ with equations in 
\citet{alan05}.
Using the IAC-Star code of \citet{aparicio04},
we then set up a series of 10~Gyr age
stellar populations corresponding to
the GSS metallicity bins in Figure~27 of \citet{ibata07}.
We normalize each bin population by the height of the bin 
(since that plot shows relative counts), and also correct
for the varying fraction of stars within the magnitude range
used in constructing the distribution.  
We assume a ``diet'' Salpeter initial mass function (defined as in 
\citealp{bell01} to have 70\% of the mass of the normal Salpeter 
IMF spanning 0.1--$100 \msun$). 
We then compute the number of counts expected within our 
color-magnitude cut chosen previously, as well as the 
stellar mass remaining in the population.
Finally, we obtain a mass conversion factor of 
$m_1 = 1.80 \times 10^4 \msun \mbox{count}^{-1}$.
%A calculation proceeding directly from the default isochrones provided 
%by the CMD web tool\footnote{{\tt stev.oapd.inaf.it/cmd}} yields 
%agreement in $m_1$ within 10\%, for both 8 and 10~Gyr age populations.
Given uncertainties in the population synthesis models, the IMF, and 
the observed stellar population of the GSS, $m_1$ is probably uncertain by 
$\tsim 0.2$ dex or so.

In our likelihood computation, we simply count the particles
within each sky bin to get $\mathbf{N_{sim}}$, 
multiply this by the mass ratio
$m_p / m_1$, where $m_p$ is the simulation particle mass,
and divide by the number of pixels in each region.
This yields the equivalent model counts per pixel in each bin,
$\mathbf{M_{\im}}$.
% = m_p \mathbf{N_{sim}} / (m_1 N_{pix})$.
Now we combine all of these quantities to obtain the image likelihood.
It is well known that using a $\chi^2$ test in the case of Poisson
statistics can lead to biased results.  However,
in our case the observed count values are large enough that we expect
this bias to be minor.  In addition, the
observational errors are actually dominated by background/foreground
fluctuations from unrelated components of M31's halo, background
galaxies, and Milky Way stars, and earlier we estimated this background
to have significant correlations between bins.  
Thus the only practical choice in this case is to use a matrix $\chi^2$ 
form for the likelihood:
\begin{equation}
\label{eqn.imlike}
L_{\im} = -\frac{1}{2} \chi_{\im}^2 = 
-\frac{1}{2} (\mathbf{D}_{\im}-\mathbf{M}_{\im}) \mathbf{C}_{\im}^{-1} 
(\mathbf{D}_{\im}-\mathbf{M}_{\im}) \; .
\end{equation}

In an earlier version of our sampling we found a sizable minority
population of states with strange-looking W Shelf morphologies, 
not matching the sharp edge visible in Figure~1.  We traced the
origin of these states to the large background uncertainties assigned
to the W Shelf bins, which in our automated procedure are influenced by
structure near M31's disk.  Fortunately, a kinematic survey along
the NW minor axis 
clearly distinguishes between the GSS debris and smooth halo
components \citep{fardal12}.  
This paper provides an accurate estimate of the fraction of red M31 RGB stars 
(those with $\feh > -0.75$, roughly corresponding to our color-magnitude cut) 
that belong to the GSS shelf: $0.84 \pm 0.05$.
This allows an independent means of subtracting the background.   
The spectroscopic survey overlays bins 10 and 12 in Figure 1, 
though with a much smaller azimuthal coverage, and 
we take this survey to be representative of these two bins,
which we combine into a single minor-axis W Shelf region.  Our estimate 
from the INT survey for the source surface density in this region was 
$2.81 \amin^{-2}$.
Subtracting the estimated Milky Way and background galaxy component of
$1.27 \amin^{-2}$, and
generously assuming 30\% error in this, we estimate the M31 and GSS 
counts together to be $(1.54 \pm 0.38) \amin^{-2}$.
Multiplying this by our estimate of the GSS fraction gives a
GSS debris count of $D_W = 1.29 \amin^{-2}$ in this region,
with uncertainty $\sigma_W = 0.32 \amin^{-2}$.
We augment the likelihood expression in equation~\ref{eqn.imlike}
with the term $L_W = -(D_W - M_W)^2 / (2 \sigma^2_W)$ to constrain
the model further.

\subsubsection{Stream and lobe position}

Our next observable is the estimated sky position of the GSS proper.
This ultimately derives from the same photometric
survey map as was used to compute the image likelihood.  
However, we find it useful
to treat these two observables separately.  The way we have set up our
GSS bins, with the broad central bin and the two outer regions added
to give a single ``outrigger'' bin, ensures the GSS position has only
a second-order effect on the surface density term in the likelihood,
making it legitimate to include both quantities.
The stream position can be estimated without doing an $N$-body
simulation using the stream-orbit approximation of \ref{sec.orbital}. 
It can also be estimated where uncertainty in the foreground/background
subtraction makes the surface density unreliable, such as the GSS at
large radius.  Conversely, the density can be measured even in regions
where a stream position is ill-defined, as in the W Shelf.

Our starting point is the location of the 8 southern fields F1 through
F8 taken with the CFHT12K camera on CFHT in \citet{alan03}, which were
centered close to the GSS transverse density peak and were used to
obtain the distances used later.  We use the slope in $\xi$ and $\eta$
of these positions to define an M31-centered coordinate system 
$(m,n)$, with unit vectors having coordinates in the $(\xi,\eta)$ system 
of $\mathbf{\hat{m}} = (0.504, -0.864)$ (along the stream to the SE) and
$\mathbf{\hat{n}} = (-0.864, -0.504)$ (across it to the SW).  
Here we treat $m$ as the independent and $n$ as the dependent variable.
We then compare these field locations with the map of ``red'' RGB stars in 
the INT survey \citep{irwin05}, which is here functionally identical with
the map shown in Figure~\ref{fig.map}.  From a histogram of counts as
a function of $n$, we find a peak at about $n=0.34 \degree$.  This
also appears consistent with the peak positions in bins centered on
the individual fields, so we use this location in each bin as our
stream position $D_{\pos}$; these are offset by about $0.07 \degree$ to the SW
from the field centers in \citet{alan03}.  We discard fields F1 and
F8, since in both fields it is difficult to estimate a well-defined
stream position.  The width of the stream obtained using
Gaussian$+$baseline fits is about $0.2\degree$ \citep{font06}, and we
conservatively take this to be the observational error $\sigma_{D,\pos}$
on the stream positions.  The positions used are shown in Figure~\ref{fig.map} 
and listed in Table~\ref{table.obsdata}. 

To minimize noise in the likelihood, and avoid contamination from
other radial wraps, we use the stream locus
estimated from the stream-orbit approximation rather than computing it
directly from the simulation.  We interpolate the value of $n$ 
from this locus at the values of $m$ corresponding to the field positions
in the table.
We tested the performance of this model
in some early samples of simulation runs, obtained in the 
manner discussed in Section~\ref{sec.samples}.
We estimated the mean position directly from the simulation
particles, applying distance and velocity cuts to isolate the stream,
and iteratively applying a smooth window function in the transverse
direction to find the mean $n$ around each field position.  From
comparison of these estimates with the stream-orbit estimate we found
on average the latter was biased by $-0.13\degree$, so we correct the
estimate by this amount in our likelihood function.  We also found a
degree of scatter, which can be represented by a statistical error on
each data point of $\sigma_{M,\pos} = 0.24\degree$.  We add this
theoretical uncertainty in quadrature to the observational error,
so that 
$\sigma_{\pos} = (\sigma_{D,\pos}^2 + \sigma_{M,\pos}^2)^{1/2} = 0.32\degree$.
The position term in the likelihood is then 
\begin{equation}
L_{\pos} = - \sum_{i=1}^{N_{\pos}} \frac{ (D_{\pos,i} - M_{\pos,i})^2 }{ 2 \sigma^2_{\pos} }
\end{equation}
%% Note: I think I deleted the directories where these corrections were obtained.
%% it was done very early, 10/08, all of my resample directories postdate that.

Since our scenario asserts the second loop of the orbit (the next one
in front of the GSS), lies in the NE Shelf region, we add another likelihood
term to encourage this behavior.  We base the term on the apocenter
position of this orbital loop, which we set to lie in the vicinity of
$\xi = 1.8\degree$, $\eta=0.65\degree$; this translates to a projected
radius $D_R = 1.9\degree$ and a position angle of $D_\PA = 70\degree$.  
We assigned loose constraints of $\sigma_R = 1.0\degree$ and
$\sigma_\PA = 25\degree$.  
The corresponding model values are taken from the apocentric
position of the NE Shelf in the orbital calculation.  The lobe likelihood term 
\begin{equation}
L_{lobe} = -\frac{ (D_R - M_R)^2 }{ 2 \sigma^2_R} 
-\frac{ (D_\PA - M_\PA)^2 }{ 2 \sigma^2_\PA}
\end{equation}
helps nudge the parameter states early on in the sampling
toward the desired orbital trajectories.  Otherwise it is not a
particularly strong constraint on our model, and later we check
that removing it makes little difference to our most important parameters.

\subsubsection{Stream distance}

Another likelihood term is based on the distance to the GSS proper, as
estimated from the tip of the red giant branch (TRGB) by
\citet{alan03}.  We use the outermost 7 GSS fields from that paper;
the innermost field does not follow the trend of the others, and it is
likely to be contaminated by other structures including M31's disk and
inner spheroid.  The positions of these fields are shown in
Figure~\ref{fig.map} and listed along with the derived distances $D_d$
in Table~\ref{table.obsdata} .  We assume a statistical distance error
in each field of $\sigma_{D,d} = 20 \kpc$ as estimated in McConnachie
et al.  To compare to simulations we assume the distance of M31 is 780
kpc, as estimated using the same TRGB method in \citet{alan05} and
\citet{conn12}.

These TRGB distances probably also have systematic offsets due to the
combination of stellar population uncertainties and the specific
algorithm for determining the tip magnitude.  We assume these
systematic offsets affect each field's distance by the same factor. 
Equivalently to
first order we can substitute a linear \textit{shift} $\Delta$ in each
field's distance, defined so that the true distances are
systematically larger by $\Delta$ than the observationally derived
values.  

We assume a Gaussian prior for $\Delta$ described by mean
$\overline{\Delta}$ and dispersion $\sigma_{\Delta}$.  \citet{brown06a}
measured the red clump brightnesses within the stream, 21~kpc in
projection from the center of M31, and in a minor-axis,
spheroid-dominated field.  They found the stream was only $(11 \pm 5)
\kpc$ more distant than the spheroid position, which is likely at 
M31's distance to within a few kpc.
Interpolating the distance at their stream field from the measurements
of \citet{alan03} gives $50 \kpc$.  Therefore we assume a best-estimate 
offset of $\overline{\Delta} = 11 - 50 = -39 \kpc$.  We take the offset's
uncertainty to be $\sigma_{\Delta} = 25 \kpc$, reflecting 
systematic uncertainty in both the TRGB and red clump distance methods.

As with the stream position, we estimate the model's stream distance
$M_d$ from the stream-orbit approximation rather than computing it directly
from the simulation.  This approximation is accurate enough given the
large uncertainties in the observed distances, and avoids the 
contamination from other radial wraps of the GSS that is possible
in unusual orientations.

The combination of the distance likelihood and the prior on 
$\Delta$ can now be written as
\begin{equation}
\label{eqn.distlikefn}
L_d = 
 \sum_{i=1}^{N_d} \frac{(D_{d,i} + \Delta - M_{d,i})^2}{\sigma_d^2}
 + \frac{(\Delta - \overline{\Delta})^2}{\sigma_{\Delta}^2} %\; \\
\end{equation}
Although in principle we could keep $\Delta$ as a parameter in our
MCMC run, we are only interested in the posterior distribution
marginalized over that nuisance parameter.  An easier way to calculate
this distribution is to solve given each choice of model distance values 
$\mathbf{M}_{d}$ for the maximum likelihood value of $\Delta$, which is
\begin{equation}
\widehat{\Delta} = 
  \left( \sum_{i=1}^{N_d} \frac{M_{d,i} - D_{d,i}}{\sigma_i^2} 
 + \frac{\overline{\Delta}}{\sigma_{\Delta}^2} \right)
  \left( \sum_{i=1}^{N_d} \frac{1}{\sigma_d^2} 
 + \frac{1}{\sigma_{\Delta}^2} \right)^{-1} \; .
\end{equation}
Using the likelihood value $L_{d}$ given
by inserting $\Delta = \widehat{\Delta}$ in Equation~\ref{eqn.distlikefn},
one can show this gives the correct posterior distribution of the other
parameters marginalized over $\Delta$.

\subsubsection{Stream velocity}
Our final likelihood term incorporates
estimates of the GSS velocity based on six Keck/DEIMOS
spectroscopic fields in the GSS core, presented in \citet{ibata04} 
and \citet{gilbert09}.
All the mean velocities are based on fits in the latter paper,
which discusses fields a3 (comprising 3 neighboring masks), 
H13s (2 masks), and f207 (1 mask) in detail.
Velocities for fields s1, s2, and s6 (1 mask each) are based on fits
to the individual stellar velocity points shown in \citet{ibata04}.
We have assigned a larger blanket uncertainty of $25 \kms$ to these
three fields, because of concerns that the formal uncertainty is an
underestimate in fields with fewer stars such as these, especially 
in the presence of a clumpy halo component.
The data values $D_{v,i}$ and uncertainties $\sigma_{D_v,i}$ are listed
in Table~\ref{table.obsdata}.
% Note: hoping that the Ibata04 velocity values and errors will be superceded
% soon by work in progress.  Early Trethewey draft showed some differences from
% Ibata04 despite being based on same data, not sure if that's
% persisted...that is part of my rationale for the large errors 
% assigned to Ibata04 points.  While new values won't get incorporated 
% in this methods paper, should have a followup as larger/better datasets
% emerge---perhaps along with using PAndAS.
When comparing to simulations, we assume M31's systemic velocity
is $-300 \kms$ \citep{devauc91}. 
We omit fields in the extended envelope to the SW,
since from inspection of our previous models 
\citetext{\citetalias{fardal07}, \citealp{fardal08}}
we found this can have significant, model-dependent offsets from the
core field values.  Figure~\ref{fig.map} shows the positions of these
fields.

The stream velocities are based on small fields where both
observations and simulations have small count numbers.  To 
mitigate the effects of simulation noise discussed in
Section~\ref{sec.sampling.noise}, we use the stream-orbit approximation to
calculate the model stream velocity $M_{v,i}$, as we did already with the 
stream position and distance.  Because the observed velocity uncertainties 
are relatively small, we measured the offsets between our
approximate treatment and the velocity values estimated from the
$N$-body simulations in several early samples of runs obtained 
with the methods in Section~\ref{sec.samples}.  We noted systematic
trends in the parameters $F_p$ and $M_{200}$, and used the results
from a linear regression in these parameters to apply slight ($\tsim 5
\kms$) corrections to the estimated velocities.  Typical remaining
offsets between estimated and simulated velocities are of order 5--$10
\kms$, and generally much less than the simulated velocity dispersion
in a given field.  The worst disagreements come in the outermost
fields in cases where the velocity distribution is the furthest from a
Gaussian, and characterizing it with a single number is difficult.  To
be on the safe side we add a rather conservative simulation rms error 
of $\sigma_{v,M} = 15 \kms$ in quadrature with the observational rms errors:
$\sigma^2_{v,i} = \sigma^2_{D_v,i}+ \sigma^2_{v,M}$.  
The statistical weights assigned to the different fields
are thus more equal than initially apparent.  
We then assume the data points are independent, and calculate the
velocity term in the likelihood using a $\chi^2$ form:
\begin{equation}
L_v = - \sum_{i=1}^{N_v} \frac{ (D_{v,i} - M_{v,i})^2 }{ 2 \sigma^2_{v,i} }
\end{equation}

We have now enumerated all the terms in the likelihood function.  The
total likelihood function is obviously quite complicated, and one may
wonder about the sensitivity of the results to the various parts of
the likelihood.  Our Bayesian treatment enables us to test this formally
once we obtain the parameter state samples in the next section.  
We simply obtain new likelihood values for the states
with the altered likelihood
function, weight each state by the ratio of new to old likelihoods, 
and compute new parameter means and dispersions for the sample
using these weights in the averaging.  
Effects we tested include: changing the
mean of the distance prior from $\overline{\Delta} = -39 \kpc$ to
$\overline{\Delta} = 0$, altering the position formalism to assume
correlated errors, omitting the NE Shelf lobe term, and dropping the
model correction terms to the position or velocity.  The resulting changes
in the means of important parameters such as $F_p$, $M_{200}$, and $M_{sat}$
are well below their dispersions, indicating the results are not
strongly sensitive to any one of these assumptions.

\subsection{Sampling method} 
\label{sec.sampling}
By now we have defined several spaces of physical parameters (such as
the satellite and halo mass, etc.), and defined both the likelihood
and the prior probability on these model spaces.  The goal of this
project is to sample the posterior distribution---the prior times the
likelihood---in parameter space, and thereby obtain the probability
distribution of the parameters and any quantity relating to them,
including the distribution of quantities obtained directly from the
simulated satellite debris.  We first discuss the main strategies we
use to perform this sampling, and then discuss the impact of
simulation noise on our results.
 
\subsubsection{Sampling strategies}
\label{sec.sampling.strategy}
We sample the posterior distribution using the Metropolis-Hastings algorithm
for Markov Chain Monte Carlo (MCMC) \citep[e.g.,][]{gelman03,press07}.  
This uses a series of steps in one or more Markov chains.  At each
step in the chain, we use a ``proposal function'' to generate a trial
parameter vector, or ``state'', from an old one.  Depending in part on the
new value of the posterior probability compared to the old one, the
chain may move to the trial state or remain at the old one (in which
case the trial state is discarded).  
After a sufficient number of steps, which is highly problem-dependent,
the chains should reach an equilibrium where it fairly reflects
the posterior distribution.
We use the Bayesian Inference Engine (BIE), a
parallel, scriptable, checkpointing MCMC code written in
%C++ 
{\hbox{C\raise.15ex\hbox{\footnotesize ++}} 
that implements numerous sampling algorithms \citep{weinberg12bie}.  
In the discussion here we assume the
posterior distribution has a simple, basically unimodal
form, which we will test later for our real problem.

The choice of proposal function in MCMC is important.  The step size
must avoid the extremes of small steps that fail to go anywhere, as
well as large steps that fail to focus on the peak of the sampled
distribution.  In addition, correlations between parameters can make
convergence difficult.  The specific MCMC technique we prefer is
Differential Evolution MCMC \citep[DE-MCMC,][]{braak06}.  This is
nearly a standard Metropolis-Hastings algorithm, except that the
proposal function depends on the population of current states in the
different chains.  The chief advantage of this technique is that it
automatically adapts the widths and covariances of the proposal
function to the shape of the sampled function, obviating the need to
set these manually.  The method generates a proposed state vector for
a given chain by incrementing the current state by the difference of
two state vectors from other chains, scaled by a parameter $\gamma$.
For a normal distribution in a parameter space of dimension $d$, the
optimal scaling parameter is $\gamma = 1.7 d^{-1/2}$.  Our posterior
distributions are not exactly normal distributions, and we find it
more effective to set the scaling parameter at a smaller value 
$\gamma = 1.0 d^{-1/2}$.  We run with 32 parallel chains, which is
enough to ensure reasonable sampling of the state difference vectors
for our dimension values.  DE-MCMC requires the chains to advance in a
synchronized fashion, which motivates a load-balancing scheme.  At
each step, we crudely estimate the computation time required for each
simulation and allocate the pool of processors accordingly, to ensure
each chain can advance to the next step in a timely fashion.

The goal of our sampling is just to make the posterior distribution
reasonably accurate, as measured for example by the quantiles of the marginal
distribution of a given parameter.
(See the discussion of the desired precision in MCMC in \citealp{press07}.)
We might choose a goal of making the 5\% and 95\% quantiles 
accurate to 5\% when compared to their difference.  
For a normal marginal distribution in a parameter $x$, 
$x_{95}-x_{5} = 3.29$, and asymptotically 
$\sigma_{5} = \sigma_{95} = 2.11 N^{-1/2}$,
implying one requires $N \approx 160$ independent samples
for $\sigma_{95} = 0.05 (x_{95}-x_5)$.
Thus only about 5 independent samples {\em per chain} 
are required to reach this precision for a normal distribution.  

To estimate how many independent samples we have, 
we estimate the ``cluster length'', the average number of states per
independent sample, as $\Sigma_l \rho(l)$, where $\rho(l)$ is the
autocorrelation of one element of the parameter vector
(normalized by the variance so that $\rho(0) = 1$) and the sum 
is taken from negative through positive values. Our cluster length
varies with different parameters and runs but is typically 
$\tsim 70$--100.  In comparison, the cluster length at the
optimal proposal step size when sampling from a $d$-dimensional 
normal distribution is roughly $3d$ \citep{gelman03}, which
would be no more than 24 for our spaces.  
The decreased efficiency in our case is due both to deviations from
a normal posterior distribution and the presence of outlier chains.
When we detect convergence, we typically sample an additional $\tsim 250$
steps or 8,000 states to ensure the distribution is adequately sampled.  
We define convergence by a combination of the Gelman-Rubin test 
\citep[see][]{gelman03} and by-eye inspection of the parameter values, 
likelihoods, and autocorrelation functions of the chains.  

Although our focus is on sampling rather than optimization, in the
course of developing the model it can be useful to precede the sampling
run with a optimization run, to obtain a single canonical model and to
check for problems with the likelihood function.  We do this simply by
making the acceptance step ``greedy'', always accepting the higher
likelihood state.  (This optimization algorithm is known as 
Differential Evolution or DE, as opposed to the sampling
algorithm DE-MCMC.)  Selection of an optimal state
is intrinsically imprecise in the face of the noise in the likelihood
function we discuss later, and this algorithm is not particularly 
quick to reach convergence.  However, it is certainly preferable
in our case to gradient-based fitting methods that assume smoothness.

In some cases, we choose our initial parameter values by sampling
uniformly from the prior.  However, for poor parameter settings that
only put tidal debris far away from the regions sampled in the
likelihood, the likelihood noise becomes very large and makes
random-walking to a better region difficult.  This suggests we should
try to initialize the chains to somewhat reasonable values.  
The ``orbital'' model uses most of the likelihood terms in
the full simulation-based models, though the star-count term 
is missing, and it is very quick to compute.  
We extract a set of parameter samples 
from the ``orbital'' sample, add random values for the satellite 
mass parameters to complete the model space, and use these values
to initialize the MCMC chains for the ``stellar'' and ``DM''
samples.  This leads to much quicker convergence for our MCMC runs 
using \nbody\ simulations.

\subsubsection{Effect of noise}
\label{sec.sampling.noise}
Our likelihood function depends in part on an \nbody\ simulation,
which stochastically samples phase space with tracer particles.  
The Poisson noise in the simulation translates to noise in the
log-likelihood function and therefore an equal noise in the
log-posterior function.  (Of course depending on the simulated
problem, particle noise can also have \textit{systematic} effects
or seed differences in outputs through chaotic dynamics.
We believe that our problem is in a simpler
regime where Poisson noise is the dominant effect.)
For simulations with too few particles, the noise may prevent the the
run from converging or render the sampling results inaccurate.
On the other hand, if many \nbody\ particles are used in an effort to
decrease the particle noise, the computation may slow down to the
point where the entire sampling project becomes impractical.
Thus it is important to determine whether there is some level 
of likelihood noise that can offer adequate sampling results.
Standard treatments of Bayesian sampling offer little guidance 
on this issue, though various approaches have been offered
in different disciplines \citep[such as][]{flury11}.
We have tried to address it ourselves with a combination
of toy analytic models and empirical tests.  

Even for a perfect model, a simulation will be offset from the true value 
by some amount, which will differ from one simulation to the next.
Let us consider for now a model in which 
the observational and simulation rms errors are Gaussian,
leading to a likelihood function with the $\chi^2$ form
(which is the indeed the form of most terms in our likelihood function).  
Suppose also the data points are uncorrelated and both the observational 
rms errors $\sigma_o$ and simulation errors $\sigma_s$ have equal values for
each data point.  To take account of the simulation errors, we must
write the log-likelihood as 
\begin{equation}
\ln \mathcal{L} = -\frac{1}{2} \chi^2 = -\frac{1}{2} \sum_{i=1}^n 
\frac{(x_{i,o} - x_{i,s})^2}{\sigma^2_o + \sigma^2_s} \; .
\end{equation}
If the simulation errors were dominant, $\sigma_s \gg \sigma_o$, 
we would find the well-known result that each
data point contributes $2$ on average to the variance of $\chi^2$,
or $1/2$ on average to the variance of $\ln \mathcal{L}$.
In general, averaging over both observational and simulation errors, 
one can show each point should contribute
$2^{-1} \sigma_s^2 (\sigma_s^2 + 2 \sigma_o^2) (\sigma_s^2 + \sigma_o^2)^{-2}$.
For the case where $\sigma_s = \sigma_o$, this implies a contribution 
to the variance of $\ln \mathcal{L}$ of $1/6$ per bin on average.
For typical MCMC calculations, the posterior sampler samples
log-likelihood values contained within 10--20 of the peak value.  In
the regime where $\sigma_s \sim \sigma_o$, one clearly cannot afford
too many terms in the likelihood before the likelihood noise swamps
the true variation in the likelihood.

What then is the effect of this noise on the parameter sampling?
We have explored the effect of likelihood noise using 
both toy analytic models and simple one-dimensional 
sampling experiments.  
We find that likelihood noise does not affect the sampled parameter
distribution as long as the noise is constant in parameter space.
If instead the likelihood noise varies strongly 
with parameter values, the estimated distribution can be highly distorted.
Another symptom of likelihood noise is that the
likelihood values are biased to high values.
In fact, the character of the Metropolis-Hastings random walk changes
to a series of infrequent jumps from one noise spike to another,
slowing the convergence to equilibrium.

One technique we have found useful is occasional resampling: 
once every $N_{rs}$ steps, we rerun the likelihood compution on 
the {\it same} parameter set, but using a new random seed,
and use the new likelihood value rather
than the old.  This dramatically cuts down on the number of states that
get stuck on noise spikes, speeding state mixing and convergence.  In
simple tests with 2-$d$ Gaussian functions, we found a wide range of
values of $N_{rs}$ were useful, adding little overhead while greatly
improving the mixing.  We adopt the value $N_{rs} = 13$ throughout.  
The drawback of the occasional resampling technique is that it can
distort the sampling results.  Because the resampled likelihood value 
is unbiased, likelihood noise enhances the probability of accepting 
a new, intrinsically inferior state in subsequent states, leading to
parameter distributions that are biased high in the tails.  For
sufficiently small noise levels ($\sigma_L \ltrsim 1$), we found this 
was not a large problem in the test runs.

From these simple tests and our experience with the GSS problem, we 
suggest these rules of thumb:
\begin{itemize}
  \item[-] The noise is usually not a major problem if it can be kept as
  low as $\sigma_L \ltrsim 0.25$--0.5 in the central regions where the
  likelihood is large.  For larger noise values, the sampling will
  encounter increasing difficulty in converging.  While there may well 
  be reasons besides likelihood noise for using large numbers of
  particles (e.g.\ to resolve physical effects such as dynamical
  friction), cutting down the noise beyond this target value will not
  greatly improve the sampling and can lead to excessive calculation times.
  Our value of 65,536 particles per simulation was chosen in accordance 
  with this guideline.
  \item[-] For moderate noise levels, occasional parameter resampling 
  as discussed above is 
  useful to reduce the stickiness of the chains and thereby speed convergence.
  \item[-]  The parameter dependence of the likelihood noise is important.
  While problem-dependent in nature, the noise will usually increase 
  for parameter settings that lead to poor average likelihood values, 
  especially when the simulations yield small numbers of particles 
  in the observed regions of physical space.  
  This can lead to ``stuck'' chains
  that fail to random-walk to better likelihood values, a tendency
  that is compounded by the nature of the DE-MCMC algorithm.  
  If the number of such chains is small, it should not affect the results.  
  If there are many such chains ($\gtrsim 10$--20\% of the total), the run 
  may need to be restarted with a larger number of simulation particles
  to reduce the noise.
  \item[-] The number of observational constraints should be kept 
  as low as possible, since each constraint adds its own noise.  
  Constraints should be kept only if they add substantial constraining 
  power to the likelihood.
  \item[-] If a term in the likelihood can be estimated analytically rather than 
  derived from a $N$-body simulation, it may be worthwhile to do so, 
  depending on how accurate the analytic estimate is and how much 
  likelihood noise is introduced by using the simulation. 
  We made extensive use of this tactic here.
\end{itemize}

\subsection{Parameter samples}
\label{sec.samples}

We have now assembled all the ingredients for our calculation, and are
ready to create a large set of parameter vectors sampled from the 
posterior distribution using the full probability model.

We first generate a parameter sample for the 7-d ``orbital'' model. This
is based only on the parts of the likelihood that can be calculated
directly from the chosen orbit with our semi-analytic technique, and
does not involve the progenitor mass.  In this calculation, we have
also omitted the informative prior on the M31 halo mass $M_{200}$, and
instead used a flat prior.   This was done principally to expand the
range of halo mass sampled, as we had early indications that the 
simulation-based sampling would prefer higher mass values than
the informative prior.   
The calculation of the orbital sample is extremely rapid because
no simulations are required.  

\begin{table*}
%\begin{minipage}{126mm}
%\begin{minipage}{164mm}
\begin{minipage}{180mm}
\caption{Parameter results}
\label{table.params}
%3-model table
\begin{tabular}{@{}c@{}rrrr@{\hspace{1.2em}}rrrr@{\hspace{1.2em}}rrrr@{}}
\hline
          & \multicolumn{4}{c}{``Orbital'' sample} & \multicolumn{4}{c}{``Stellar'' sample} & \multicolumn{4}{c}{``DM'' sample} \\
Parameter & Mean  &  Sigma & \multicolumn{2}{c}{95\% range} & Mean  &  Sigma & \multicolumn{2}{c}{95\% range} & Mean  &  Sigma & \multicolumn{2}{c}{95\% range}\\
\hline
$X     			           $&$   0.8564  $&$      1.911  $&$     -3.308  $&$      3.764   $&$    1.51  $&$      1.17    $&$   -0.77   $&$     3.84  $&$  0.90  $&$     1.12  $&$    -1.03   $&$    3.49 $\\
$Z     			           $&$    16.33  $&$      5.596  $&$      7.001  $&$      28.72   $&$    19.73  $&$     2.716   $&$    13.66  $&$     24.47 $&$   18.33  $&$     3.234  $&$     10.08   $&$    24.25 $\\
$V_x     		      	   $&$   -94.76  $&$      18.05  $&$     -133.2  $&$     -62.33   $&$    -83.9  $&$     11.2    $&$   -108.9  $&$     -64.9 $&$  -87.7  $&$     13.3  $&$    -118.6   $&$   -65.6 $\\
$V_y     		      	   $&$    183.9  $&$      33.94  $&$      124.6  $&$      255.2   $&$    173.7  $&$     22.36   $&$    137.7  $&$     226.0 $&$   177.5  $&$     25.58  $&$     136.1   $&$    240.1 $\\
$V_z     		      	   $&$   -236.2  $&$      30.60  $&$     -293.5  $&$     -178.4   $&$   -244.0  $&$     11.24   $&$   -263.8  $&$    -218.9 $&$  -247.6  $&$     16.9  $&$    -277.1   $&$   -207.8 $\\
$F_p     		      	   $&$    1.416  $&$     0.3270  $&$     0.8498  $&$      1.966   $&$    1.241  $&$   0.062   $&$    1.112    $&$     1.388 $&$   1.26   $&$    0.16  $&$    0.96      $&$    1.71 $\\
$\log_{10} (M_{200} / M_{\sun})   $&$    12.32  $&$     0.1458  $&$      11.97  $&$      12.50   $&$    12.26  $&$   0.084   $&$    12.10  $&$     12.42   $&$   12.27  $&$    0.10  $&$     12.07   $&$    12.48 $\\
$\log_{10} (M_{sat} / M_{\sun})    $&$   NA 	 $&$ 	NA	 $&$ 	NA	 $&$  NA    $&$    9.550  $&$   0.088   $&$    9.398  $&$     9.742   $&$   9.87  $&$    0.31  $&$     9.22   $&$    10.39 $\\
$\log_{10} (M_{lum} / M_{sat})     $&$    NA	 $&$ 	NA	 $&$ 	NA	 $&$  NA    $&$    NA       $&$        NA     $&$    NA       $&$      NA     $&$ -0.28  $&$    0.28  $&$   -0.71   $&$   0.35 $\\
\hline
\end{tabular}
\end{minipage}
\end{table*}

We then run our full simulation sampling technique on the 8-d
``stellar'' parameter space, generating an \nbody\ simulation for each
sample and computing the prior and likelihood functions as described above.  
This run converges to reasonable values (1.2--1.5 depending on the quantity
measured) of the Gelman-Rubin $\hat{R}$ convergence parameter \citep{gelman03}
within 1046 steps or 33,472 $N$-body simulations.
The computation took about two weeks using 256 Opteron processors in parallel.
We discard an initial burn-in period of 500 steps.  Four of
the chains remain stuck throughout this run in regions far from the
mode, with posterior values far below the maximum.  We have culled all
chains that never come within 4 of the maximum log-likelihood before
computing the parameter statistics.  
%Note: crude justification: 
%alog(0.05) = -4.60517
%2*4 = 8  (delta chisq)
%nominally 6 independent steps
%alog(1.-chisqr_pdf(7.,8)) * 6 = -3.73465
%alog(1.-chisqr_pdf(8.,8)) * 6 = -5.01559
%still worried about truncating tails of distribution though
%[Note: these Gelman-Rubin values are not great.  Possibly want to repeat 
%run with more states and 2x number particles for smaller fraction
%of stuck chains.]  
This leaves 20,888 remaining states in 28 chains.
Using the cluster length computed as
described above, we estimate that each chain contain about 6
independent steps, for about 200 independent states total.  The
frequency of jumping to a different state is 23\%, close to the
value of 25\% that is optimal for sampling from a Gaussian peak.

\begin{figure*}
\includegraphics[width=16cm]{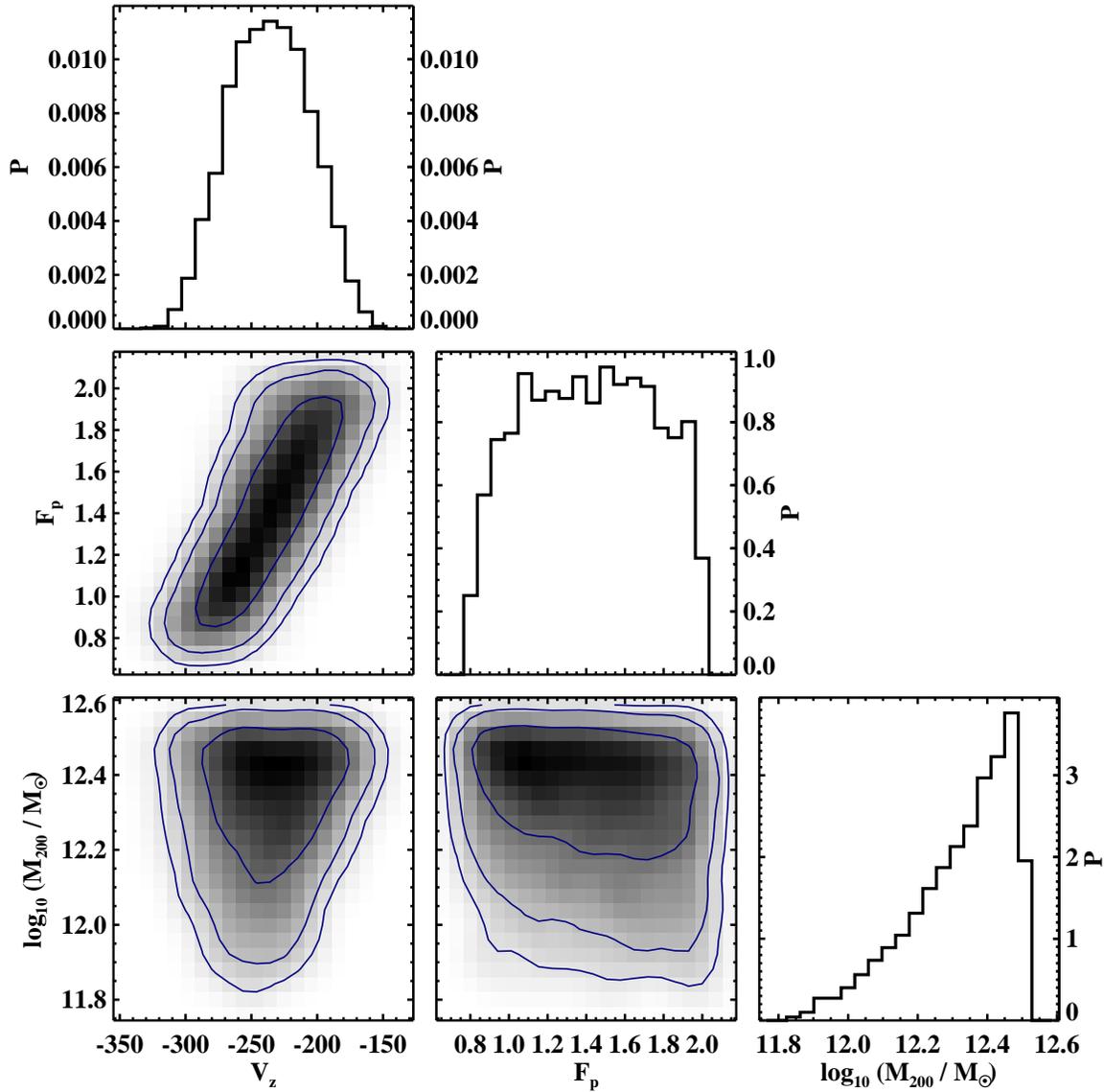}
\caption{
\label{fig.contours_orbits}
Parameter distributions in the orbital sample.  
Diagonal panels show the marginalized distributions of individual 
parameters, while
off-diagonal panels show the joint distribution of
pairs of parameters.
The parameters shown here are the $Z$ velocity at the 
orbital reference point, the progenitor phase 
$F_p$, and the M31 virial mass $M_{200}$.
The phase $F_p$ is calculated using the progenitor's
original orbital trajectory, whether or not it survives to the 
present time.
In this sample a flat prior for $M_{200}$ is used rather than 
the informative prior (see discussion in the text).
Contours enclose 67\%, 95\%, and 99\% of the probability.
}
\end{figure*}

\begin{figure*}
\leavevmode \epsfysize=14cm \epsfbox{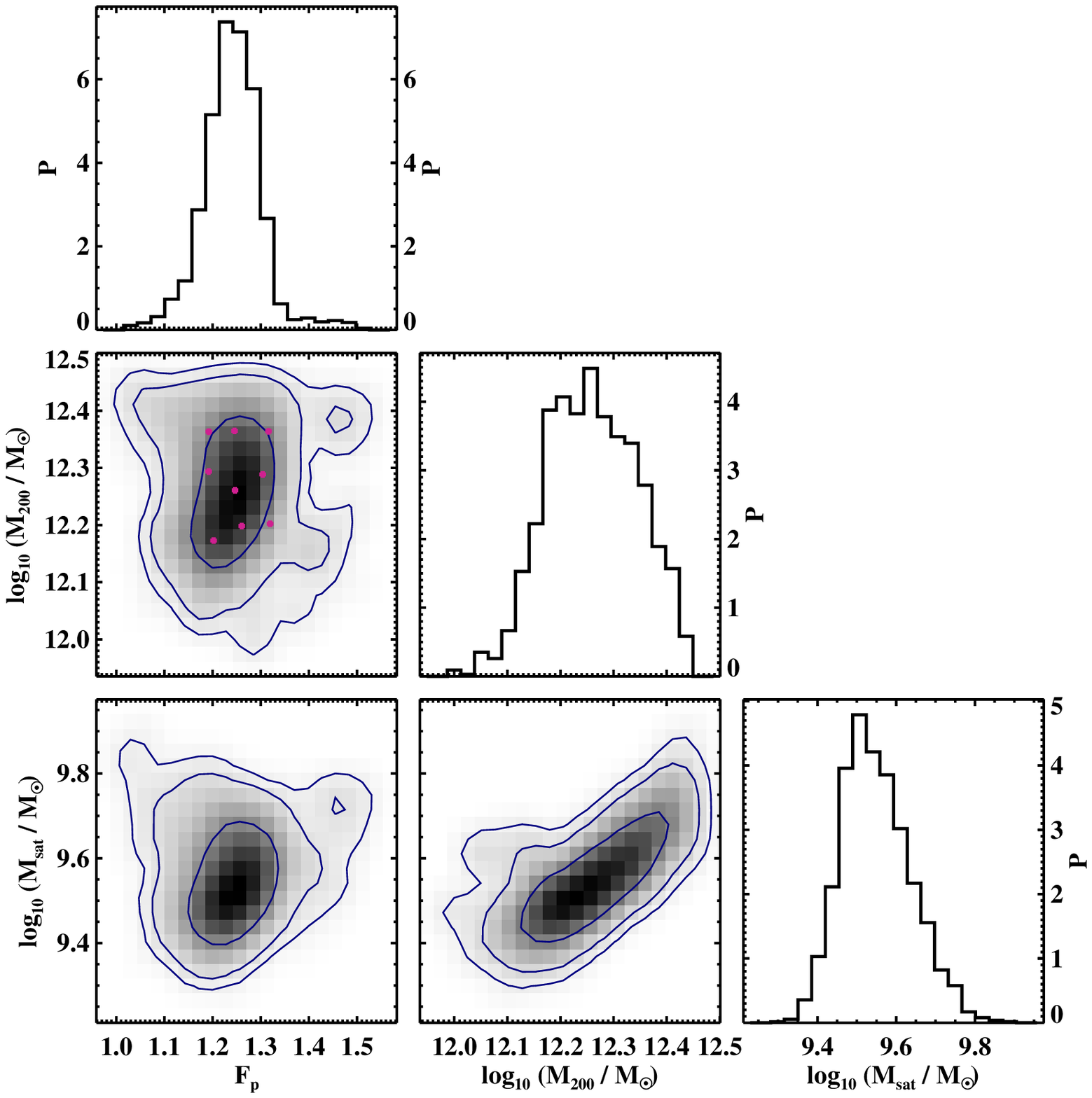} 
            \epsfysize=1cm \epsfbox[350 -470 420 -400]{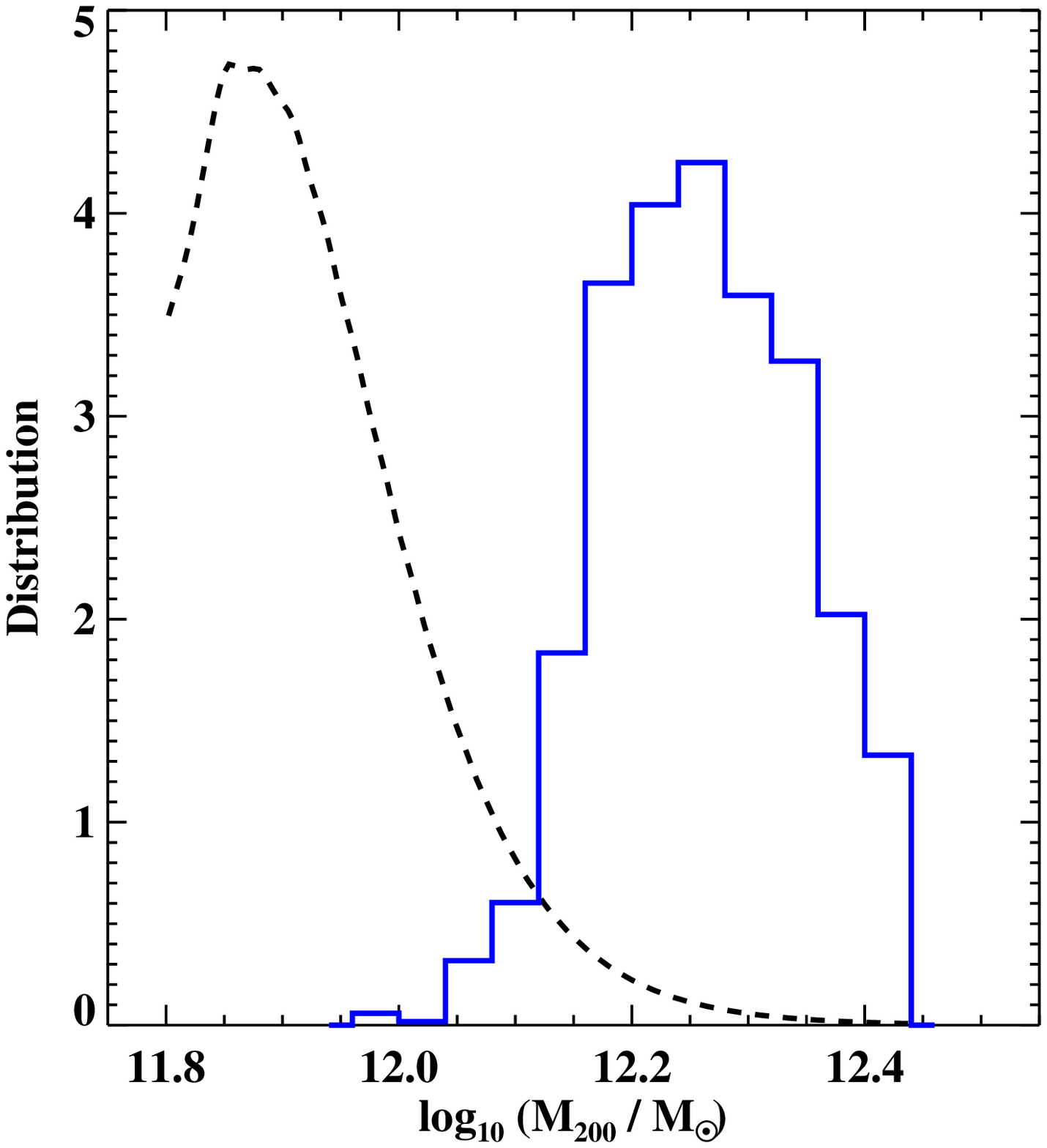}
\caption{
\label{fig.contours_st}
Parameter distributions in the ``stellar'' simulation sample.  
Diagonal panels show the distributions of individual 
parameters, while
off-diagonal panels show the joint distribution of
pairs of parameters.
The parameters shown here are the progenitor phase 
$F_p$, the M31 virial mass $M_{200}$, and the 
GSS progenitor mass $M_{sat}$. 
Crosses in the $F_p$-$M_{200}$ panel show the rough grid of states
to be used in Figure~\ref{fig.morphology}. 
Upper right panel: final distribution of the halo mass (histogram) over
the entire prior range, with prior (dashed) also shown.
}
\end{figure*}
 
Next, we conduct a run with the 9-d ``DM'' model, which includes the
effect of dark matter by the simple expedient of allowing the ratio of
luminous to total mass $M_{lum}/M_{sat}$ to vary  (implemented by 
scaling our conversion factor for RGB number to dynamical mass).  
The run converges sufficiently after 1240 steps, and we 
discard the first 500 steps and 3 outlier chains leaving
a sample of 21,460 states total, or roughly 200 independent states. 

As mentioned earlier, we also tried two other parameter spaces with 
fewer dimensions.
In our ``reduced'' parameter space, the only free orbital parameter
is $F_p$ and the remaining orbital parameters are frozen out by
fitting the orbital model as a function of $F_p$.
The ``density'' space adds freedom in choosing the progenitor's
central density.
Convergence for these lower-dimensional spaces was significantly
easier, as the chains found a good likelihood region more quickly and
took far fewer steps per independent state (the cluster lengths
were $\tsim 15$).  We therefore ran for only 453 and 657 steps 
respectively.

\section{RESULTS}  %3
\label{sec.results}
\subsection{Parameter distributions and likelihood tests}
\label{sec.params}
In Figure~\ref{fig.contours_orbits}, we show the distribution of several
parameters in the 7-$d$ ``orbital'' sample.  Here the panels on the diagonal 
show the marginal distribution of each parameter, while the others show the
joint distributions of pairs of parameters.  It is clear that the
orbital sample can place almost no interesting constraints on the
progenitor phase or M31's halo mass.  This is because a larger progenitor
phase can be traded off against a smaller orbital energy to produce
nearly the same GSS properties.  
(To be fair, the sample does favor larger values of the halo mass, and 
in this sample we have not included the informative prior on this parameter
which favors lower values.  Combining the two would yield a posterior 
distribution similar to the halo mass prior, but shifted by 0.1 dex 
to higher masses.)
The parameter results are also summarized in Table~\ref{table.params}.

\begin{figure*}
\includegraphics[width=16cm]{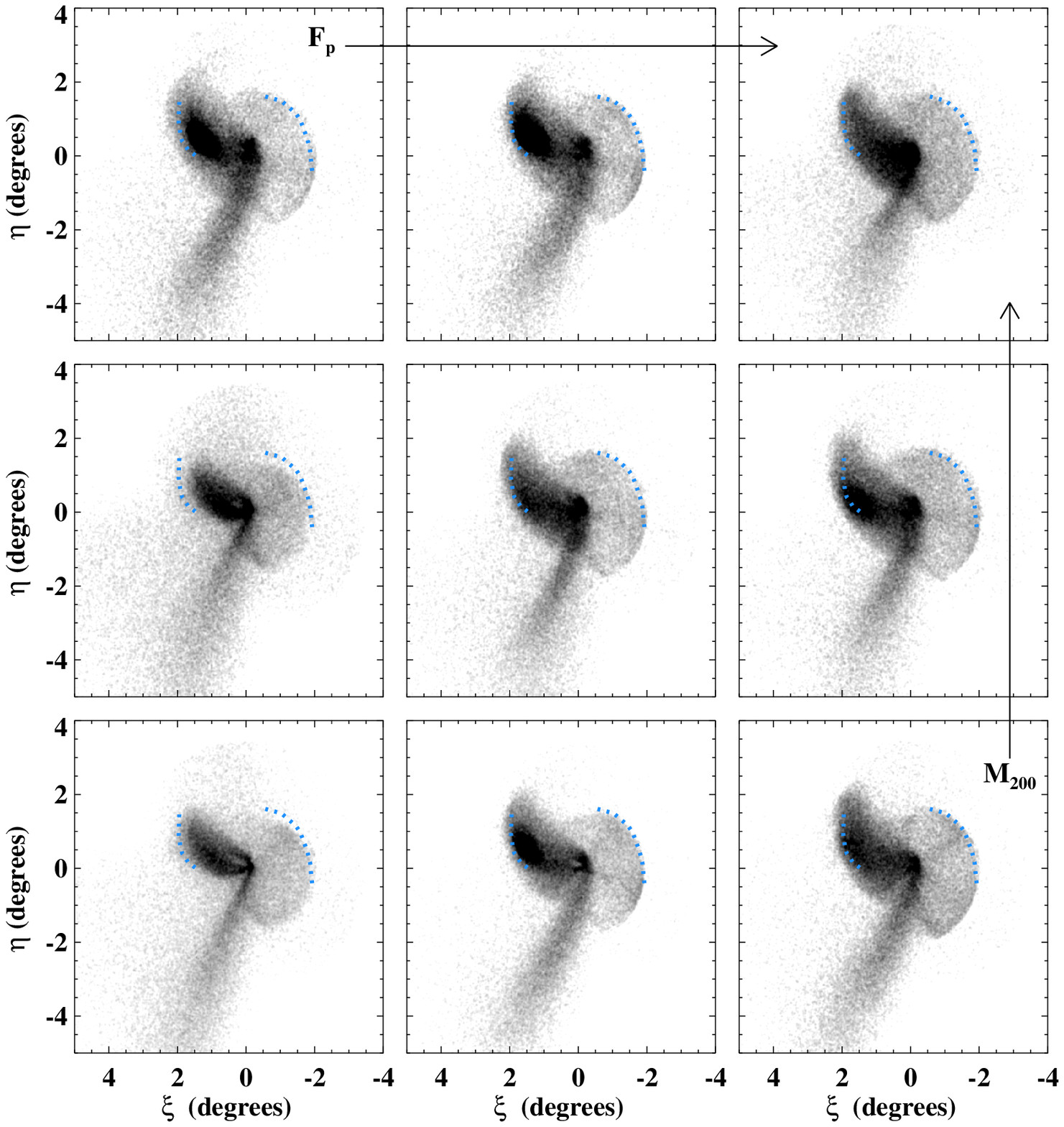}
\caption{
\label{fig.morphology}
Sky distribution for nine parameter states taken from the ``stellar''
sample, on a square-root intensity scale.  
The values of $F_p$ and $M_{200}$ for these states are
arranged roughly in a grid around the distribution mean, with
separation of about $1\sigma$ (see Table~\ref{table.params}).  $F_p$
increases to the right and $M_{200}$ increases upwards in this grid of
plots.  Dashed lines show the locations of NE and W shelves from 
\citet{fardal07}.}
\end{figure*}

In Figure~\ref{fig.contours_st}, we show the parameter distributions
for the 8-d ``stellar'' sample, which are also summarized in
Table~\ref{table.params}.  Clearly, the parameters are now much better
constrained than for the orbital sample.  In \citetalias{fardal07} we
argued that increasing the orbital phase would decrease the density in
the GSS and enhance that in the W Shelf.  In test runs where we leave
out one or the other of these constraints, we see that this is indeed
true.  Combining constraints from these two regions thus leads to an
startlingly accurate determination of the orbital phase, $F_p = 1.24
\pm 0.06$.  This then constrains the spatial pattern so accurately
that the satellite's (purely stellar) mass can also be measured
precisely, $\log_{10} (M_{sat}/\msun) = 9.55 \pm 0.09$.  In fact, the
dominant uncertainties are probably systematic, for example error in the 
mass normalization constant $m_1$ calculated earlier.  The halo mass
$M_{200}$ is the lone parameter to take on a large range of values
compared to our initial prior range, $\log_{10} (M_{200}/\msun) = 12.32 \pm
0.09$, but even it is significantly altered relative to our prior
distribution, as seen in the inset.  
Halo masses as low as our initial expectations are
incapable of generating the required GSS velocities, for our
well-constrained satellite mass and orbital phase.  We will discuss
this result in more detail in Section~\ref{sec.discussion.halomass}.  The
figure also shows a significant correlation between $M_{200}$ and the
progenitor satellite's mass.

We create a ``library'' of nearly independent simulations of this model by
rerunning one state every 100 steps for the converged part of each
chain.  This yields 
$6 \mbox{~steps} \times 28 \mbox{~chains}= 168$ states, counting only 
those in the retained chains.  The majority of these states share a 
morphology similar to that in the observed map in
Figure~\ref{fig.map}, although there are a few unusual-looking states
at low likelihood values.  We select a sample for display by
defining a grid in $F_p$ and $M_{200}$
and selecting representative states near each grid point
(see Figure~\ref{fig.contours_st} for their actual locations).
Figure~\ref{fig.morphology} displays the sky pattern of these states.
Each of the runs shows well-defined GSS, NE Shelf, and W Shelf
features.  The good agreement of the models with the observed W Shelf
edge is achieved by use of the surface density constraints 
in the W Shelf region (Figure~\ref{fig.map}).  
The states here also agree well with the
NE Shelf edge, though the variation in this shelf location is larger
in the total sample.  Some systematic variation of the morphology with
the parameter values is apparent: as $F_p$ increases from left to
right, so do the W Shelf surface density and areal coverage.  Also, as
$M_{200}$ increases from bottom to top, the GSS appears shorter and
broader.  This suggests the correlation between $M_{200}$ and
$M_{sat}$ arises because a larger halo mass speeds the return of the
GSS material, thus advancing it in phase, which then requires
a larger mass satellite to produce a GSS density consistent with
observations.  However, there are also variations not simply explained
by these two variables, including large differences in how much of a
central core or clump survives in the NE Shelf region.

\begin{figure*}
\includegraphics[width=16cm]{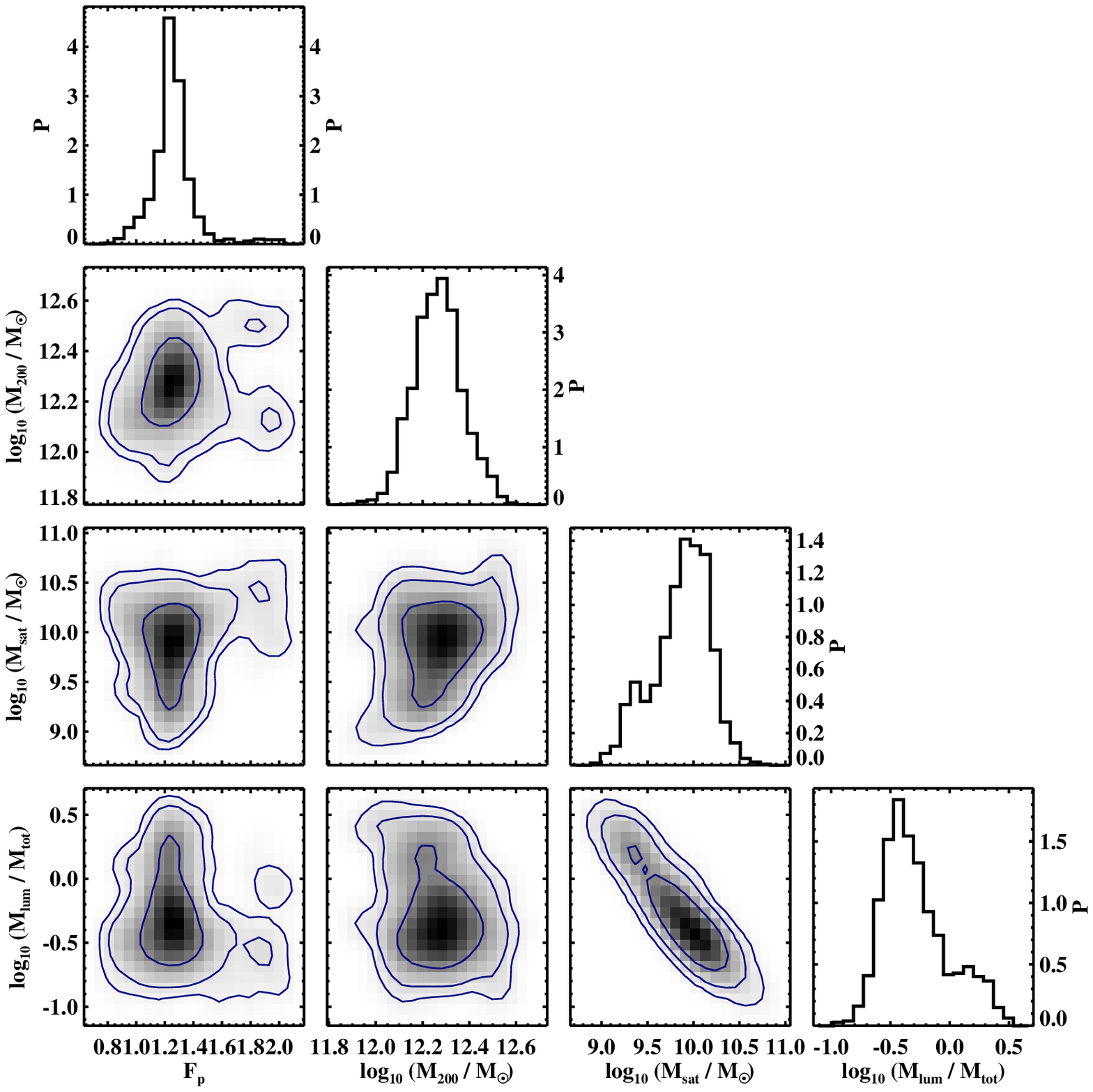}
\caption{
\label{fig.contours_dm}
Parameter distributions in the ``DM'' simulation sample.  Diagonal
panels show the distributions of individual parameters, while
off-diagonal panels show the joint distribution of pairs of
parameters.  The parameters shown here are the progenitor phase $F_p$,
the M31 virial mass $M_{200}$, the GSS progenitor mass $M_{sat}$, and
the GSS luminous to total mass ratio $M_{lum}/M_{sat}$.  
}
\end{figure*}
 
\begin{figure*}
\leavevmode \epsfysize=8cm \epsfbox{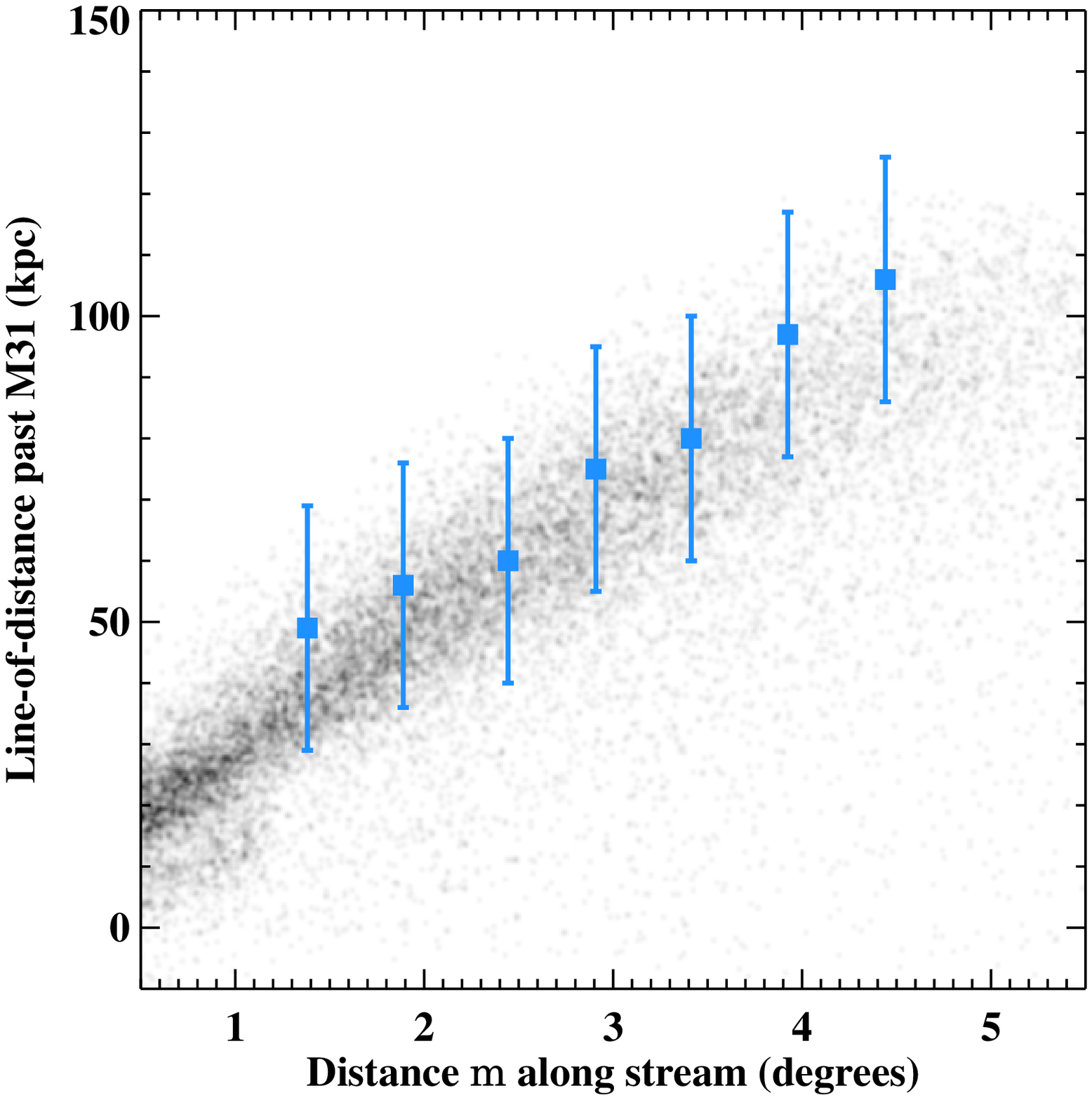} 
            \epsfysize=8cm \epsfbox{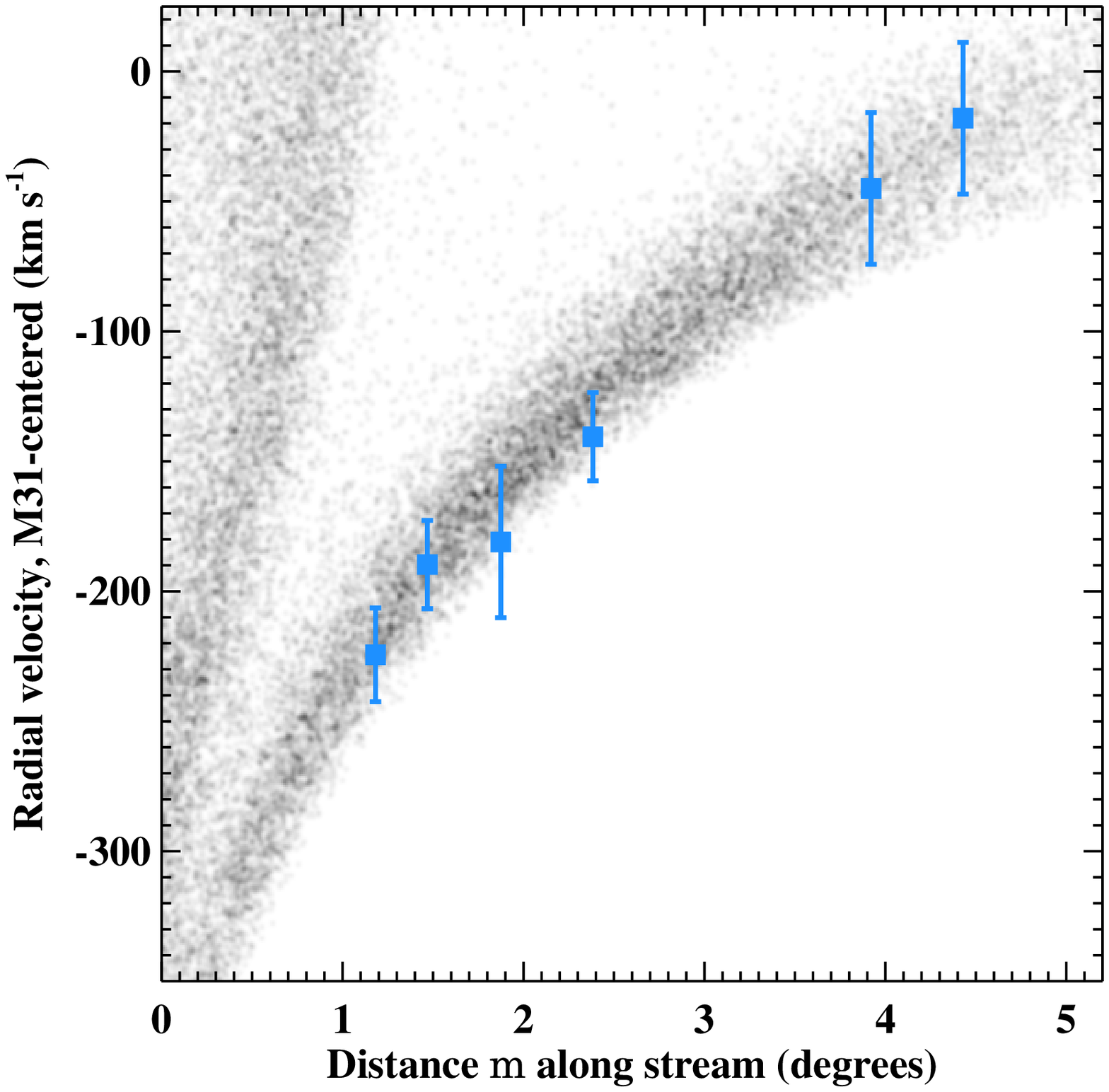}
\caption{
\label{fig.gssobs}
GSS properties of the central panel in Figure~\ref{fig.morphology}.
Left panel: distance to points along GSS, as a function of
pathlength along the stream, compared to data in \citet{alan03}.  
We apply cuts in stream transverse position $1<n<3$ and in M31-centered
velocity $v_{M31} < 30$ to select predominantly true GSS particles.
The $m$-$n$ coordinate system is depicted in Figure~\ref{fig.map}.
Right panel: mean velocity at points along GSS,
compared to data in \citet{gilbert09}.  
The triangular feature above the GSS in the simulation
is mostly NE Shelf material on the near side of M31.
}
\end{figure*}
 
In each of these resimulations we calculate the orbital
trajectory of the progenitor assuming a test-particle orbit.  
In this sample the progenitor has apocenter 
$55.5 \pm 4.5 \kpc$,
pericenter 
$2.86 \pm 0.81 \kpc$, 
orbital period 
$610 \pm 60 \Myr$, 
and it reached the disruptive pericenter 
$760 \pm 50 \Myr$ 
ago.
A few orbital quantities show strong correlations with particular input
parameters. As expected, the apocenter and orbital period are very
well correlated and both anticorrelate strongly with $F_p$.
Also, the pericenter is positively correlated with
$M_{200}$.  Mostly, however, the dispersion results from combinations
of the ``uninteresting'' orbital parameters.
In any case the results show very little dispersion in 
each quantity, showing that the model predicts a very definite
trajectory and history of interaction with M31.

The results of the 9-d ``DM'' sample 
are shown in Figure~\ref{fig.contours_dm} and 
again summarized in Table~\ref{table.params}.  The
uncertainty on the orbital phase more than doubles, $F_p = 1.26 \pm
0.16$, and that on the progenitor mass more than triples, $\log_{10}
(M_{sat}/\msun) = 9.87 \pm 0.31$.  However, the amount of luminous matter 
and its uncertainty are almost the same as before: 
$\log_{10} (M_{lum}/\msun) = 9.58 \pm 0.09$.  
$M_{lum}$ correlates with $M_{sat}$, roughly as $M_{lum}
\propto M_{sat}^{1/4}$.  Thus the amount of dark matter is highly
uncertain, but the median model contains nearly equal masses of dark 
matter and stars. The halo mass and its uncertainty are barely affected 
by the change in model.

We again create a library of nearly independent simulations by
rerunning one state every 100 steps for the converged part of each
``DM'' chain, yielding $8 \times 29 = 232$ states.  Inspection of this sample
helps us interpret the correlation of $M_{lum}$ with $M_{sat}$: higher
values of $M_{sat}$ spread out the tidal debris over a larger region,
requiring a larger $M_{lum}$ to keep the surface densities at observed
levels.  The variety in this simulation library is enhanced relative to the
``stellar'' sample, but as before most simulations share the same basic
morphology.

Most of the observational constraints are very well satisfied in both
of our models.  As an example we display the GSS distance and velocity 
for one of the ``stellar'' model states in Figure~\ref{fig.gssobs}.  The 
observed gradients along the stream---infalling from behind M31, and
speeding up along the way---are very well reproduced in our model.

The maximum likelihood value in the ``stellar'' sample is
$-9.4$, corresponding to an effective $\chi^2$ value of 18.9 for 28
degrees of freedom (6 stream positions, 7 stream distances, 6 stream
velocities, 2 NE lobe positions, 1 galaxy potential prior, and 14
surface density regions, minus 8 parameters), which is quite
reasonable.  (Simulation noise should bias the maximum
likelihood value upwards.  Estimating this effect with
a mean noise of $\sigma_{LF} = 0.5$ and 8 degrees of freedom
suggests the 99.8\% quantile of the resimulated distribution
is a good estimate of the true maximum LF value, and we
have in fact defined our maximum likelihood value this way.
We have also incorporated the galaxy prior value into this
estimate of $\chi^2$, via a term $-2(\ln[P(M_h)] - \max\{\ln[P(M_h)]\})$,
since it is variable whereas all other parameters have uniform priors.)
%To do this we simply add  -2.*(prior_{gal} - max(prior_{gal} to chisq...
%corrected these for max gal lf, which is -6.1413934 in stellar run
%  and -6.1856744 in ml run - close enough to use same value
% max lf value computed from lfstats2.pro is -14.96
If anything, the fit is unexpectedly good, with a probability
of a better one by chance of only 10\%.
Though it is possible we have underestimated the positive bias of
simulation noise on the maximum $\chi^2$, we ascribe the low
$\chi^2$ value primarily to some combination of luck and
possible overestimation of the observational errors.
The maximum likelihood value in the ``DM'' model is $-9.0$,
corresponding to an improvement in $\chi^2$ of only 0.8
achieved with 1 extra parameter, which indicates no
superiority of this model.

Besides the halo mass parameter, the only
datapoint that is consistently more than $1\sigma$ off is an outer
W~shelf region; this region's background-subtracted surface density
is negative, which the simulations naturally cannot reproduce.
A common though mild 
deficiency of the models is in the GSS region, where the models
find it difficult to achieve a density gradient along the stream as
large as in the observations.  It remains to be seen how much of this
discrepancy is due to unsubtracted variations in the background, and
how much reveals true deficiencies in the model.  The density
distribution within the satellite may play a role, perhaps motivating
a more complex model than the Plummer sphere employed here.  
The models also may introduce more curvature into the stream velocity than 
suggested by the observations (as seen in Figure~\ref{fig.gssobs}),
though usually only a single data point (at $2.3 \degree$) is visibly
offset.  The scatter between different resimulated models is
comparable to the size of the observational error bars.

We can test whether either of our models is preferred by means of the
Bayes ratio, which is the ratio of the integrated likelihood or
``Bayesian evidence'' in each model.  General use of the integrated
likelihood would require us to take more care in specifying our prior.
But in our case, where the models are ``nested'' and one simply
incorporates an additional parameter, the uncertainties in specifying
the prior mostly cancel out in the Bayes ratio.  The integrated
likelihood is not straightforward to compute from a MCMC parameter
space sample, and the ``harmonic mean'' formula sometimes recommended
for this purpose yields spectacularly poor results here.  
We calculate the integrated likelihood by means of the volume tesselation 
method of \citet{weinberg12evidence}.
The maximum log-likelihood is only 0.25 larger in
the ``DM'' than in the ``stellar'' model, and this is canceled out by
the smaller fraction of parameter space occupied, giving a final Bayes
ratio close to unity.  Another way to compare the
models is the value of $-2 \ln L = \chi^2$ averaged over the parameter
sample \citep{johnson05}.  This is again very close in the two models.  
Therefore, a modest dark matter contribution in the progenitor
is quite plausible, though not in any way preferred by the data, 
whereas a cosmological baryon to dark matter ratio is ruled out.

We note that the use of an informative prior on $M_{200}$ is important
in obtaining its nicely peaked distribution visible in
Figures~\ref{fig.contours_st} and \ref{fig.contours_dm}.
Specifically, the prior factor from \citet{geehan06}, based on
kinematic observations of M31 and its halo tracers, is peaked around
$M_{200} = 12.0$ and acts to pull down the distribution on the high
side (see the separate panel in Figure~\ref{fig.contours_st}).  
Without this factor the peak would probably be raised by $\tsim
0.2$ dex.  The likelihood function from fitting the GSS, in contrast,
prefers higher masses and pulls down the low side.  The other factors
in our halo prior are less important and can be significantly altered
without much effect on the results.  Thus any follow-on from this
study should probably begin by reassessing the prior from M31's
kinematic tracers.  In contrast, the other priors used in this work
are mostly not very significant.
For example, multiplying the flat prior on $\log_{10} M_{sat}$
in the ``stellar'' model by $M_{sat}^{-1}$ shifts the distribution by
less than 0.02 dex, and the same is true for $\log_{10} M_{stellar}$
in the ``DM'' model.  The choice of prior gives a larger offset for
$M_{sat}$ in the DM model, simply because this parameter is not
very well constrained.

We mostly used the ``reduced'' parameter sample as a preliminary step
towards the larger spaces investigated earlier.  We also used this
sample to compare to the ``density'' sample, where the characteristic
satellite density $\rho_{sat}$ is allowed to vary.
The resulting parameter sample shows essentially the same results for
$F_p$, $M_{200}$, and $M_{sat}$ as for the ``reduced'' sample.
This indicates the satellite density has little effect on
the model for most of the parameter range.  However, there is a tail to
high values of $M_{sat}$ and $\rho_{sat}$.  Inspecting simulations
with these values, we find they share a large intact core that retains
much of the original mass, thus requiring a larger mass to populate
the GSS and W Shelf regions with debris.  These runs look to be
inconsistent with the lack of an obvious progenitor in the NE
shelf region, and we expect that this tail could be ruled out by
detailed observations in that region.

\subsection{Model testing and predictions}
\label{sec.results.tests}
Beside the reasonable fit obtained by our parameter samples, is there
other evidence to believe in our general scenario?  We can look for this
in a variety of quantities that were not included in the likelihood
function, because they were difficult to measure robustly or because
of concerns over contamination from other structures in M31.  We can
also offer some predictions for quantities that have not yet been
observed.

\begin{figure}
\includegraphics[width=8cm]{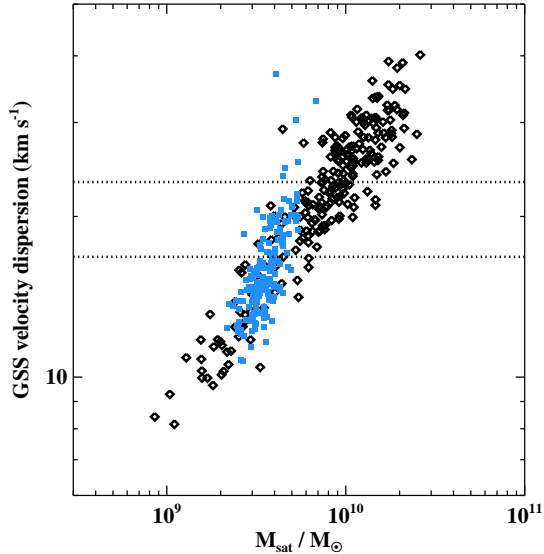}
\caption{
\label{fig.vdisp}
GSS velocity dispersion for the resimulated states in the ``stellar''
sample, plotted as a function of progenitor mass $M_{sat}$.
Open diamonds show the ``DM'' sample, and 
small filled squares show the ``stellar'' sample.
The calculation of the dispersion is discussed in the text.
Dashed lines show the minimum and maximum estimated values
for the dispersion in the three on-stream fields from 
\citet{gilbert09}.  
}
\end{figure}

One piece of evidence is the generally good agreement of our model
states with the observed morphology of red RGB stars around M31
(compare Figures~\ref{fig.map} and \ref{fig.morphology}).  Recall that
only selected regions of this map were used in the fitting procedure,
although we also included a term loosely constraining the NE Shelf
lobe.  In a randomly selected collision model it would be quite easy
to produce jets or clouds of metal-rich debris in various directions 
which are inconsistent with observations; these unwanted features are 
generally avoided in our simulated states.  

We omitted the velocity dispersion in the GSS from our likelihood
function for several reasons.  First, the dispersion is not entirely
straightforward to measure, either in the simulations or observations;
in some cases the GSS velocities can have a distribution that is far
from Gaussian (often asymmetric to the positive-velocity side), and
details of model fitting can affect the results.  Also, we have
assumed a hot, spherical progenitor; while prior experience with
colder progenitors \citep{fardal08} has not shown dramatic effects on
the velocity dispersion, in principle a cold progenitor can reduce the
dispersion.  It is nevertheless worth comparing the dispersion to the
observations.  For each state in the ``DM'' sample, we estimate the
velocity dispersion around each of the fields f207, H13s, and a3,
which are assigned velocity dispersions of $23.2^{+7.2}_{-5.0}$,
$21.3^{+4.0}_{-3.2}$, and $16.8^{+4.6}_{-3.3} \kms$ in
\citet{gilbert09}.  For each field center we first estimate the median
velocity, weight each particle with a window function, 
then estimate the mean, recompute the weight, and re-estimate
the mean.  
(We used a window with a flat top of width $80 \kms$ 
and a rounded falloff of $40 \kms$ on each side to 
gradually downweight outliers from the stream, but 
the window shape is not very important.)
We use the mean of the three field center measurements as
our velocity dispersion measurement for each resimulated state.  As
seen in Figure~\ref{fig.vdisp}, this turns out to be a function of
$M_{sat}$ with a small scatter for most states.  

The observed range of velocity dispersion estimates from
\citet{gilbert09} is shown by the dotted lines.  This suggests that
both the extreme high and low values of $M_{sat}$ in our DM sample are
unrealistic, with the caveats noted before, whereas values of 3--$10
\times 10^9 \msun$ are favored.  We can formalize this sense through a
posterior predictive check \citep{gelman03}, where we draw samples
from the model velocity dispersions and fold in the observational
errors to generate mock replicants of the data.  The observed average
data velocity dispersion of $20 \kms$ lies at the 78\% quantile of the
replicants using the ``stellar'' sample, and the 39\% quantile using
the ``DM'' sample.  Neither value is particularly extreme, which
bolsters the sense that our model is reasonable.

\begin{figure}
\includegraphics[width=8cm]{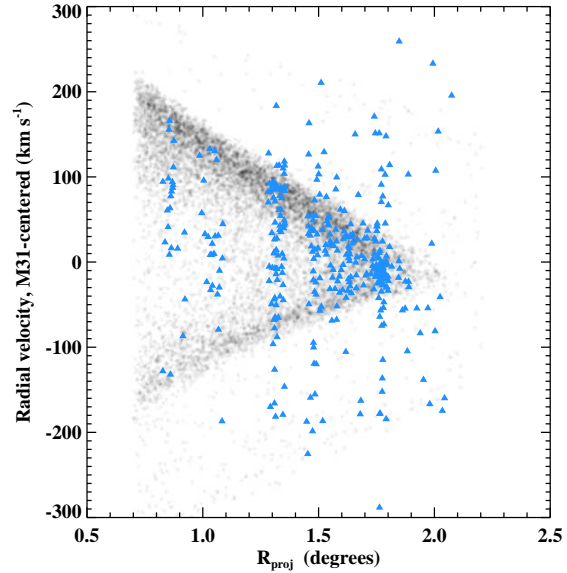}
\caption{
\label{fig.wshelf}
Velocity as a function of projected radius, 
for stars on the NW minor axis of M31.  
Here we average together all nine states in 
Figure~\ref{fig.morphology} to produce the grayscale map,
as they are all reasonably similar.
Simulation particles are selected in the range
$|X_{M31}| < 0.3\degree$, $0.7\degree < Y_{M31} < 2.2\degree$, where 
$X_{M31}$ and $Y_{M31}$ increase along M31's SW major and NE minor
axes respectively.  Triangles indicate the stars classified as M31
giants in the sample of \citet{fardal12}; these observed stars include
not only shelf material but any other components at this position.}
\end{figure}

We made no explicit use of the W Shelf kinematics in fitting the
observed data.  \citet{fardal12} found the velocities along the NW
minor axis, when plotted as a function of radius, could be well described
as a mixture of a hot spheroid and the wedge pattern expected from a
radial shell \citep{merrifield98}.
Inspecting the kinematics in this region for the states shown in
Figure~\ref{fig.morphology}, we find no major differences, so we show
these combined into one distribution in Figure~\ref{fig.wshelf},
together with the observed M31 stars from \citet{fardal12}.  (See that
paper for further details of the observational sample.)  The agreement
of the simulations with regions of enhanced density is remarkable.
Inspecting our entire sample of states, we do find some variation in
aspects such as the radius of the wedge and the constrast between
wedge edges and interior.  However, the overall morphology is quite
robust, again bolstering our model.

\begin{figure}
\includegraphics[width=8cm]{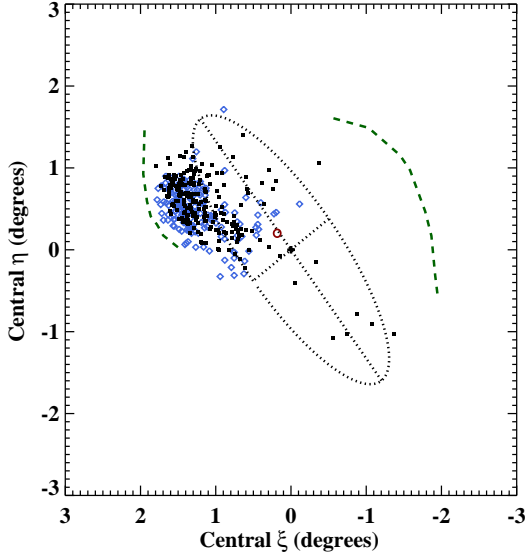}
\caption{
\label{fig.objpos}
Location of the progenitor's central material on the sky 
at the present day.  The sample uses the resimulated states
for the ``stellar'' sample (diamonds) and ``DM'' sample (squares).  
The definition of the central
location is discussed in the text.  Annotations are the same as 
Figure~\ref{fig.morphology}.  The circle near M31's center
shows the location of the overdensity found by \citet{davidge12blob}.
}
\end{figure}

\citet{gilbert07} found a wedge-like cold feature on the SE minor axis
of M31 terminating at a projected distance of about 
$R_{proj} = 1.3\degree$, and suggested
this was the fourth wrap of the GSS predicted by \citetalias{fardal07}.  In
the states resampled from both ``stellar'' and ``DM'' models, it is
quite common to find such a feature, but its strength and definition
are highly variable.  Often it is completely absent, either because the
shell itself is absent or because it does not overlap the SE minor
axis.  The shell tends to be stronger for larger $M_{sat}$, so the
occurence of a ``SE shelf'' feature is more common in the ``DM''
model.  In general the shell extends in an arc around much of 
M31, which suggests future observations might detect such a feature
at $R_{proj} = 1.3\degree$ at different locations.  While the shell is
clearly sensitive to the input parameters, indicating it might be a
strong constraint on models, we also suspect its properties would be
strongly influenced by rotation of the progenitor or a triaxial
potential.  Measurement of the shelf surface density in the
\citet{gilbert07} sample would also be required to make a quantitative
comparison with the simulations.  Thus at this point neither
observations nor models are at a stage where we can use this feature
to constrain the models, but it might become very useful in future work.

For our ensemble of model states, where is the debris from the central
core of the satellite located, and should it be clearly evident?  The
actual state of this central debris can vary from tightly bound to
highly dispersed, as can be seen in Figure~\ref{fig.morphology}.  We
determine the location of this debris by selecting the 100
lowest-energy particles in the initial progenitor for each resimulated
state, then measuring their mean present-day sky position.  The
results are shown in Figure~\ref{fig.objpos}.  Except for a few
outliers, which generally have quite low likelihood values, the central
debris lies in the NE Shelf region, in many cases projected directly
against the disk.  The low surface densities relative to M31's disk
may make it difficult to determine the core location from imaging
alone.  The core debris velocity can take on a large range of values, $-49
\pm 67 \kms$ in our ``stellar'' sample, so while this is usually distinct
from M31 disk velocities (which are $\tsim 80 \pm 40 \kms$ in these
core positions) it cannot be relied on to uniquely identify the GSS
core.  It is worth noting that none of our central cores, or even
central orbits, pass near M32 or NGC 205 when considered in position
and radial velocity space.

Recently \citet{davidge12blob} image-processed 2MASS data to discover
an overdensity in M31's disk, at about 3.5 kpc from M31's center.  He
found a magnitude $M_K = 6.5$ (corresponding to a stellar mass of
$\tsim 3 \times 10^8 \msun$) and a size of about 1 kpc, and a
composition rich in AGB stars.  The size and mass of this overdensity
make it a plausible candidate for the GSS core.  By comparison to
Figure~\ref{fig.objpos}, it can be seen that the Davidge overdensity
lies at the small-radius edge of the progenitor locations in our
sample of states.  The nearby progenitor cores in this sample are
outgoing rather than infalling, with M31-centered line-of-sight
velocities varying from $-300 \kms$ to $-100 \kms$ depending mostly on
their orbital phase.  The disk on this side has positive velocity
relative to M31 center, so spectroscopy of stars in this region could
yield a diagnosis of this object's nature.  Another possibility 
{\it not} covered by our model is that dynamical friction has slowed
the core down substantially and caused it to fall in much faster than
the rest of the NE Shelf material, so that it would be infalling again
at the present time.  The sheer amount of material liberated from the
GSS suggests that dynamical friction is unlikely to be so effective at
the last pericentric passage, due to the progenitor's lowered mass,
and we regard this scenario as unlikely.

\begin{figure*}
\includegraphics[width=16cm]{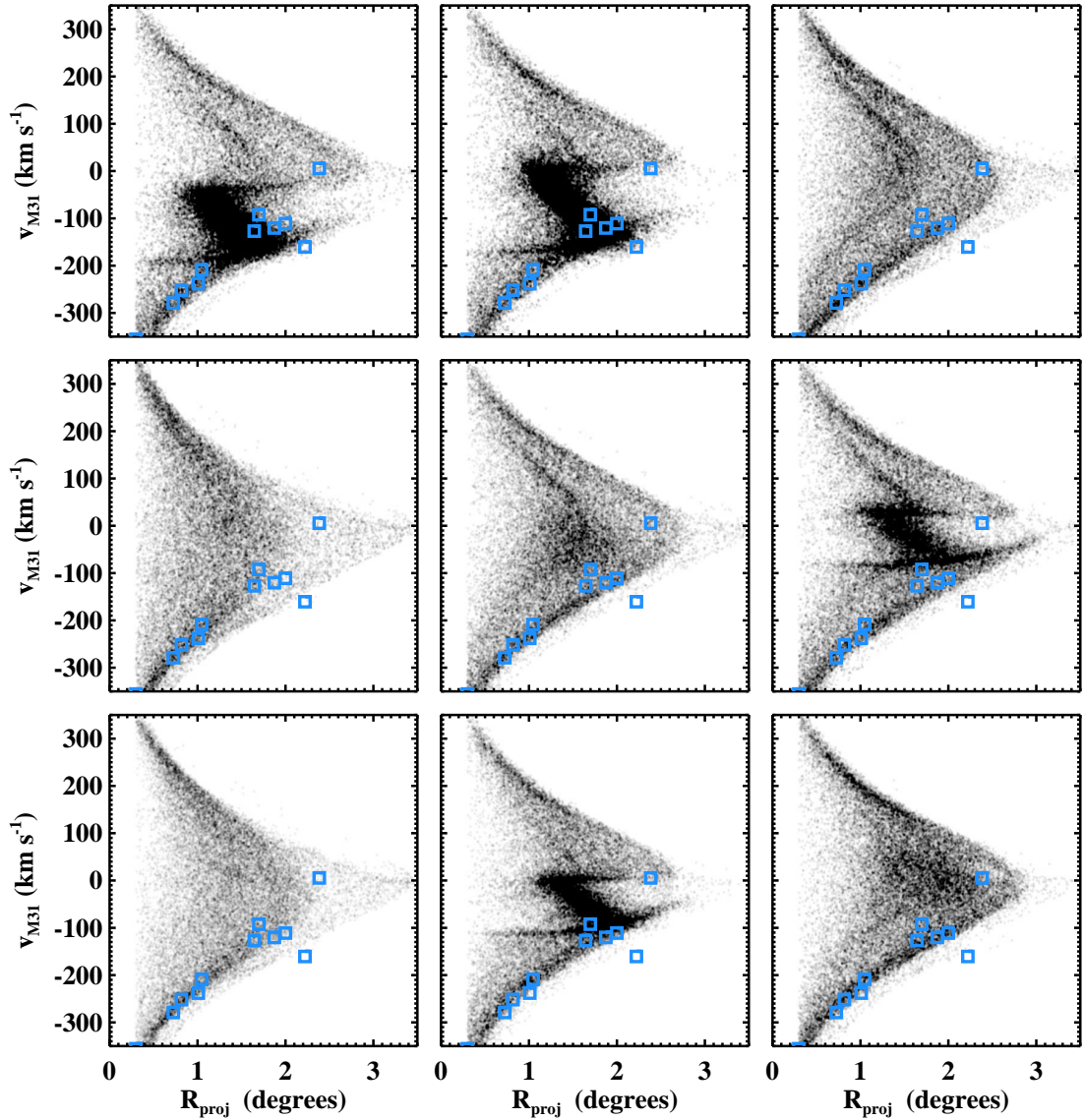}
\caption{
\label{fig.neshelf}
Velocity as a function of projected radius, for stars in the NE Shelf
region on and beyond M31's NE side.  The plots show the same nine
states as in Figure~\ref{fig.morphology}.  
Particles for this plot are selected in the region 
$X_{M31} < -0.2\degree$ and $|Y_{M31}| < 1.5\degree$, where 
$X_{M31}$ and $Y_{M31}$ run along M31's SW major and NE minor axes respectively.
The squares show the planetary nebulae labeled as ``Stream'' or 
``Stream?'' from the NE Shelf region in the sample of \citet{merrett06}.}
\end{figure*}

Detailed inspection of the NE Shelf kinematic tracers will probably be
necessary to isolate whatever remains of the GSS progenitor core.  In
Figure~\ref{fig.neshelf}, we show the velocity versus projected radius of
simulation particles in the NE Shelf region, for the same states as in
Figure~\ref{fig.morphology}.  All states display an overall ``wedge''
morphology, usually with a strong concentration at the lower boundary
which is formed by a caustic in velocity space.  As previously seen in
\citetalias{fardal07}, this caustic appears to match the planetary nebula
``stream'' detected by \citet{merrett06}.  This caustic was not fitted
to the planetary nebula data, but rather forms naturally as a product of the
debris geometry and gravitational potential in our simulations.

While the general pattern of the NE Shelf kinematics is similar in all 
models displayed in Figure~\ref{fig.neshelf}, there are also 
significant variations.  Some models have strong concentrations at
the wedge boundaries, while other spread their stars more evenly
throughout the wedge.  In most models the wedge shape is warped from a
pure triangular shape, as a result of second-order effects discussed
in regard to the W Shelf region in \citet{fardal12}.  In addition to
the lower caustic many models show an upper one, which has not yet been
detected.  The most striking effects result from the presence of an
intact core at the last pericentric passage, resulting in a
characteristic zigzag-shaped clump in these plots.  
Finally, some models show the
presence of other radial shells, resulting in a second wedge visible
within the main one.  Clearly it is difficult to make too definite
predictions about the NE Shelf kinematics, but by the same token these
measurements could be quite useful in constraining the GSS model and
locating the core debris.
%Plot NE shelf PNe sample, in $v_r$ vs $R$?

In summary, the models agree well with several observables that were
not explicitly part of the fitting process: the sky distribution of
metal-rich RGB stars outside the fitting regions, the kinematics in
the W and NE Shelf regions, and the velocity dispersion in the GSS
itself.  The latter quantity is only satisfied for total progenitor
masses in the middle of our current distributions, pointing to a
possible further constraint on the model, with caveats noted earlier.  
The model also makes some fairly robust predictions.
The debris from the core of the satellite should be located in the NE
shelf region, most likely somewhere in a circle of radius $0.6
\degree$, and with velocities more negative than expected for the disk
in the same location.  The NE Shelf should feature stars at positive
as well as negative velocities, possibly including a caustic mirroring
the one tentatively detected at negative velocities.  Finally, there
are features that are highly variable, including the total mass and
areal extent of the shell with radius $1.3 \degree$ corresponding to
the fourth orbital wrap, the detailed kinematic pattern in the NE
Shelf region, and the velocity and concentration of the core debris.
These variable features may be the key to reducing the allowed
parameter space and allowing yet tighter constraints on M31's halo
mass.

\section{DISCUSSION} %6
\label{sec.discussion}
\subsection{Bayesian Simulation Sampling} 
\label{sec.discussion.bayes}
The results in the previous section show the power of a Bayesian
formalism combined with $N$-body models to constrain physical
parameters in a complex dynamical situation.  We refer to this as
Bayesian simulation sampling.  To our knowledge, this paper
represents the first application of this technique to models of
merging galaxies.
The generic concept of using simulations in a Bayesian likelihood
evaluation has however been applied in other disciplines 
\citep[e.g.,][]{flury11}.
Our results show it is possible to obtain well-specified parameter
estimates, uncertainties, and covariances for merger models, even in cases
where simulations are necessary to estimate the observables.  This
demonstration itself may be more important than the problem-specific
results presented here.

Previous sections commented on some pitfalls of the Bayesian
simulation sampling method.
The main ones are the large computer time required,
and biases and convergence difficulties
stemming from the stochastic nature in the simulation.  
We have already discussed some of the techniques used to speed up
the calculation, including:
\begin{itemize}
\item[-] use of DE-MCMC to automatically choose the shape and scale of
  the proposal function;
\item[-] resampling of the likelihood values, to reduce the effects of noise
in the likelihood;
\item[-] assigment of reasonable initial chain values based on simpler orbit 
  calculations, instead of randomly sampling from the prior;
\item[-] use of a load-balancing scheme which assigns different numbers
  of processors to different likelihood evaluations, depending on
  their computational difficulty.
\end{itemize}

We anticipate there are numerous other ways in which our method 
could be improved.  These might include:
\begin{itemize}
\item[-] treatment of likelihood noise through averaging or smoothing.   
When we sample the same point in parameter space 
we might be better off averaging the values obtained for that point,
rather than simply replacing the value as we
have done.  It might also be possible to smooth the likelihood over 
neighboring samples of parameter space 
(see the model emulation technique of \citealp{bower10}).
\item[-] Some directions in parameter space are much easier
to explore than others.  For example, the parameter
$M_{lum} / M_{sat}$ that controls the mass-to-light ratio
is trivial to change and does not require a new $N$-body calculation.
Also, the potential is exactly symmetric about
the galactic $z$ axis and is roughly symmetric about the other 
axes, given our assumed spherical halo potential.  We could 
alternate Gibbs sampling along these ``special'' directions 
with Metropolis-Hastings sampling along ``generic'' 
directions \citep{gelman03}.
\item[-] Breaking our likelihood function into an easy part (based
on the orbital calculation) and a hard part (based on a simulation).
The sampling can then be broken into one Metropolis-Hastings step 
based on the easy part, and one performed only afterwards on the hard part.  
Rejected steps in the easy part would free up processors that can  
assigned to calculations in the hard part using our load-balancing 
scheme.
\end{itemize}
Depending on the problem, we expect the combined effect of these
techniques could speed up convergence by a significant factor.

Of course, if the structure of a stream can be described adequately
by analytical methods instead of $N$-body simulations (which is not
the case here), a Bayesian approach becomes much easier.  Many papers
on tidal streams simply assume the central path of the stream follows
the orbit of the progenitor.  Correctly relating the orbit of a
progenitor to the central locus of a stream is an ongoing research
problem, and several approximations besides that employed here have
been proposed \citep{johnston98,varghese11,eyre11}.  Which
approximation is best in a given case may depend on the orbit of the
progenitor, its degree of intactness, and the observables at hand.
Bayesian analyses of tidal streams using any of these assumptions
include those of \citet{koposov10} and \citet{varghese11}.

Our own problem was made easy in certain respects by the reasonably
simple structure of our posterior function, which had a single
dominant mode.  Other attempts to fit mergers or tidal substructure
with simulation sampling may lead to much more complicated likelihood
and posterior functions, especially when fewer types of observational
data are available or when the oscillatory behavior typical of
dynamical problems leads to multiple modes.  In many such cases, the
problem may be unfeasible given the large investment of computer time
required by the Bayesian simulation sampling technique.  In other
cases, one may be able to isolate several well-separated modes
representing alternative scenarios, impose artificial constraints on
the model restricting it to a particular mode, then apply simulation
sampling to explore the structure and implications of the domain
around each mode.  It may be argued that our particular assumptions
about the nature of the observed ``shelf'' morphology and direction of
motion of the GSS progenitor are restrictions of this kind.  As
demonstrated in Section~\ref{sec.results.tests}, an advantage of
Bayesian simulation sampling is the easy production of large
simulation libraries with parameters sampled according to the
posterior distribution.  These enable clear observational predictions
which can help confirm or rule out the choice of a particular mode.

\subsection{The GSS Model}
\label{sec.discussion.gss}
The model for the GSS structure here is that of a satellite disrupted
essentially in a single pass, resulting in the trailing GSS as well as
a number of other orbital wraps.  While the basic model cannot yet be
considered proven, it does pass a number of observational tests, based
on both morphological and kinematic data.  Our model implies that 
the GSS is due to a previously unknown satellite of M31, not any of the
currently intact ones such as M32 or NGC 205.

The general pattern of the individual simulations is similar to that
found in earlier work.  The simulation used in \citet{fardal12} is one
fairly representative element of our ``stellar'' model space, and was
in fact obtained during an earlier iteration of our Bayesian sampling.
The simulation of \citetalias{fardal07} is obtained with a slightly
different form of the galaxy potential, but is otherwise similar to a
state within our current model.  However, the stream velocities in
this simulation are too high by about $50 \kms$ on average, a
deficiency remedied by the larger values of $M_{200}$ in the current
samples.  \citet{mori08} used a live M31 model in three simulations
that otherwise followed the general pattern of the \citetalias{fardal07} 
model.  Without direct constraints from the star-count map, these
models and the earlier ones of \citet{fardal06} used a much larger
trial range of physical parameters such as $M_{sat}$ than shown to be
allowed in this paper.

The results in Section~\ref{sec.results} assign the GSS progenitor a
stellar mass of $(3.7 \pm 0.7) \times 10^9 M_{\sun}$.  This mass is
just about equal to that of the LMC, according to the model of
\citet{vdmarel02}.  Thus the GSS progenitor was either the fourth or
fifth most massive galaxy in the Local Group as recently as 1~Gyr ago.
Using simple estimates for a 10 Gyr stellar population, we find a $V$
luminosity about a factor of 3 lower than the LMC, due to a higher
$M/L$ than expected for the actively star-forming LMC population.

%% Note: from pegase 2.0 dir, salpeter and salpeter_cont
%% V mag:
%% burst, 1.0 msun
%% 7.063 8 Gyr 
%% 7.228 10 Gyr burst
%% cont formation, 1e-3 msun/Gyr = 1.e-12 msun/yr:
%% 10.659 8 Gyr 
%% 10.605 10 Gyr
%% formula for comparing same stellar mass:
%% L_cont/L_burst =  10.**(0.4*(V_burst-V_cont)) /(1.e-3 * time in yr)
%% using 8 gyr numbers gives ratio = 
%% IDL> print, 10.^(0.4*(7.063 - 10.659)) / (1.e-3 * 8.)
%%       4.55523
%% factor of 3 stated in text seems reasonable

Is our stellar mass reasonable?  We can check this by means of the
stellar metallicity in the GSS, estimated in several datasets to be in
the range $\feh = -0.7$ to $-0.5$ within the GSS
\citep{raja06,brown06b,ibata07,gilbert09}.  Using a solar metal
fraction of $Z_{\sun} = 0.019$, this translates to $\log_{10} Z
\approx -2.3$, which is well within the trend of stellar metallicities
shown by local galaxies at our derived mass \citep{dekel03,woo08}.

The dark matter in the progenitor is less well constrained by our
modeling.  Limits on the heating of M31's disk place an upper limit on
the GSS progenitor's mass of about $5 \times 10^9 \msun$
\citep{mori08}.  In addition, the stream velocity dispersion
constraints in Figure~\ref{fig.vdisp} independently suggest a mass of
$\ltrsim 10^{10} \msun$.  Therefore we have less than a factor of two
room for dark matter mass in the progenitor at the time of disruption.
This is not as implausible as it may seem.  A baryonic/dark mass ratio
of $\ltrsim 2$ within the main body of the galaxy is reasonable for
galaxies of this mass \citep[e.g.,][]{vdmarel02}.  
We infer that the bulk of the original dark matter associated with the
progenitor, most of which must have extended beyond its stellar body,
was stripped off in previous orbits.

The picture of multiple encounters with M31 agrees with the orbits
derived from our modeling, which place the apocenter well within M31's
virial radius.  Several orbital passes are probably necessary to lose
enough energy and angular momentum to shrink the orbit and allow a final
disruptive pericentric passage.  It is possible that interactions 
with massive third bodies such as M33 may play a role as well.

According to our model, the disruptive pericentric passage 
that formed the GSS took place
$760 \pm 50 \Myr$ ago, small in cosmological terms.  This is much more
recent than the last clear period of star formation ($\tsim 4$--5~Gyr)
found from modeling of HST/ACS data on the GSS down to the main
sequence in \citet{brown06b}.  We ascribe this to an earlier
ram-pressure stripping of the GSS's interstellar medium, which would
imply a hot halo in M31.  
%Note : ram pressure typically removes the outer
%gas before fully stripping the gas, and it is possible that star
%formation continued in the progenitor's core to a much later date.
%According to our model, this core should appear somewhere against the
%NE part of M31's disk or in the NE Shelf region, and may or may not be
%characterized by a large overdensity.  
%[It would be interesting to
%search for a clump of blue stars with atypical kinematics?]
On the other hand, the model puts this disruptive encounter much too
far in the past to induce the expanding star-forming wave that
\citet{block06} envisioned to explain the 10~kpc ring of star formation 
in M31.  That model fairly reliably requires a collision 210 Myr ago,
since the outward propagation speed of the star-forming wave in this
model is closely tied to M31's accurately known rotation curve.

Our model still has certain deficiencies, notably the 
transverse distribution in the stream.
We have previously argued that the transverse distribution is best explained 
with a rotating progenitor \citep{fardal08}.  When it comes to a complex
phenomenon such as the GSS merger, of course, it is likely
that all models be will wrong on some level; the question is 
whether the disagreement indicates an uninteresting discrepancy,
an interesting direction in which the model could be improved,
or a serious flaw indicating the model is fundamentally wrong.
The generally good agreement with a complex set of observational
data indicates to us the model is generally plausible at this point.
Any alternative model for the stream should be compared to data 
at a similar level of detail before it can be considered viable.
Forthcoming wider-field data from the PAndAS survey can be used to 
refine the observational input, in part by better assessing
the background contamination model.
That survey may also show signs of a more complex initial structure
than our simple Plummer model, such as remnants of a disk which
may form cold arc-like features, or remnants of a halo which might
have been stripped off in earlier encounters with M31.
New spectroscopic samples of RGB stars, when compared to 
our libraries of fairly sampled simulations, can also be
used to test the model further in the near future.

\subsection{M31's halo mass}
\label{sec.discussion.halomass}
Our models find the halo mass to be 
$\log_{10} (M_{200}/\msun) = 12.27 \pm 0.10$.
Translating this to a threshold commonly used to represent
the virial density, 100 times the critical or about 400 times 
the background density, gives $\log_{10} (M_{100}/\msun) = 12.33 \pm 0.10$,
or $M_{100} = (2.1 \pm 0.5) \mtwelve$.

In the following discussion, for consistency with most of the
cosmological literature, we translate to a ``virial'' mass defined 
by $M_{100}$ as opposed to $M_{200}$.  We also use approximate
translations between mass measures suggested by our
observationally-constrained mass model of M31.
(See \citealt{vdmarel12} for a discussion that covers some 
of the same issues, but assigns different emphasis to various
methods of measuring mass.)

Earlier work has exhibited tension between observational measures of
M31's stellar and halo mass on the one hand, and expectations from the
general galactic population and Local Group-based dynamical arguments
on the other.  For example, using a set of dynamical tracers in
the outskirts of M31, \citet{evans00b} found a total mass of 
0.7--$1.0 \mtwelve$.

The models of \citet{widrow03} that use an NFW halo imply a virial 
mass $M_{100} = 0.9$--$1.5 \mtwelve$.  
Using orbital fits to the GSS, \citet{ibata04} found a 
mass within 125 kpc of only $0.75_{-0.13}^{+0.25} \mtwelve$,
suggesting a virial mass of $(1.0 \pm 0.5) \mtwelve$.
\citet{seigar08} found a virial mass of $0.82 \mtwelve$ 
by fitting M31's rotation curve with a model of adiabatic contraction.
\citet{geehan06} inferred a virial mass of $0.77 \mtwelve$.
\citet{watkins10} used tracer mass formulae and 
data on the dSph population to estimate M31's mass within 300~kpc as 
$(0.85 \pm 0.24) \mtwelve$ to 
$(1.6 \pm 0.4) \mtwelve$, depending on 
which galaxies were assumed to be virialized and other assumptions.
These values suggest virial masses of 
$(0.77 \pm 0.25) \mtwelve$ to $(1.6 \pm 0.5) \mtwelve$.
There are several uncertainties not incorporated in the 
Watkins estimates, for example the effect of the satellite 
density slope (taken to be $-2.1$ in Watkins but 
estimated as about $-1.0$ in \citealp{richardson11}).
%Note: this would scale the halo mass by a factor 0.6?
%Also: haven't accounted for the large errors in the measured distances?
%Or: how valid are their assumptions that everything is a power law?
%Or: can you measure the mass out to r=300 kpc with satellite when the
%surveys mostly probe to R=150 kpc?  
A similar calculation using globular clusters instead of dSph
yielded $(1.2 \pm 0.2) \mtwelve$ to $(1.5 \pm 0.2) \mtwelve$ within 200~kpc,
depending on the slope of the assumed potential \citet{jovan13}.
It is likely that most of the above studies have failed to include some
important systematics in the error estimate.  We also note that 
several methods are drawing upon similar datasets, for example those
that use satellites as tracers, so the estimates are not independent.

Estimates of the stellar mass in the bulge and disk
include $9.5 \times 10^{10} \msun$ \citep{widrow03}, 
$10.4 \times 10^{10} \msun$ \citep{geehan06}, 
$11.4 \times 10^{10} \msun$ \citep{seigar08},
and $10.1 \times 10^{10} \msun$ \citep{tamm10},
%1.07e11 msun \citep{gauthier06}
This implies that stars constitute about half of M31's halo baryons,
with estimates ranging as high as 86\% \citep{geehan06},
if we assume a virial mass $10^{12} \msun$ and 
a cosmic baryon to total matter ratio $\Omega_b / \Omega_m \approx 0.15$.

These high stellar fractions conflict with values inferred from halo
abundance matching and related techniques, where surveys of the
general galactic population are compared with the expected halo and
subhalo mass distribution in an LCDM cosmology.  
These methods suggest the stars of central halo galaxies take up only 
10--20\% of the halo baryons on average, 
\citep[e.g.][]{guo10,behroozi10}, even at the halo mass
corresponding to peak efficiency.  For a M31 stellar mass of 
$0.10 \mtwelve$, in line the estimates given earlier, the formula
in \citet{guo10} converted to virial mass gives an M31 virial mass of
about $7.4 \times 10^{12} \msun$, far above the observational
estimates of M31 halo masses.

This tension echoes a current puzzle about the Milky Way, which also
by some measures has a low estimated halo mass for its stellar mass.  
For example, using halo star velocities \citet{xue08} 
estimate a virial mass $(1.0 \pm 0.3) \mtwelve$.  
\citet{smith07} estimate the local escape velocity and thereby a virial 
mass of $1.4^{+1.14}_{-0.54} \mtwelve$.  
In contrast, a halo mass of $2.0 \mtwelve$ would be most appropriate
for a stellar mass of about $5.5 \times 10^{10} \msun$ \citep{guo10}.
It would be very strange to have both large Local
Group galaxies lie far off the typical stellar-halo mass trend.  Also,
the classical timing argument for the local group \citep{kahn59}
suggests the M31 and the MW virial masses combine to 
$5.2 \mtwelve$ \citep{li08} 
to $(4.9 \pm 1.6) \mtwelve$ \citep{vdmarel12}, about twice the sum 
of the typical observational estimates given earlier (although
the uncertainty in this mass is large and that of M31 as derived
from this method is even larger).   

We now reconsider the problem, using our new, higher estimate of M31's
halo mass.  First we re-estimate the baryonic mass of M31, using the
value of $L_K = 13.7 \times 10^{10}$ derived from the Spitzer $3.6\micron$
luminosity and estimated $K-\mbox{[3.6]}$ color in
\citet{barmby06}.  Using the typical value of $B-R \approx 1.5$ as in
Barmby et al, we find $M/L_K = 0.62$ based on \citet{bell03b} 
using their stated correction term of $-0.15$ dex to translate to a 
\citet{kroupa01} initial mass function.  
This yields a stellar mass of $8.6 \times 10^{10} \msun$ for M31.

We can only expect the M31 and MW estimates to be consistent with halo
abundance matching if there is some scatter in the relationship
between halo mass and stellar mass.  Various arguments lead to a
scatter of $\sigma_\ast \tsim 0.1$ dex, which is explicitly modeled in
\citet{yang09} and \citet{behroozi10} among others.  The virial masses
yielding a median stellar mass of $8.6 \times 10^{10} \msun$ are $5.0
\mtwelve$ and $7.6 \mtwelve$ respectively using results from these two
papers.

However, when inverting the relationship to get the median halo mass
at a given stellar mass, we should take account of the halo luminosity
function, which lowers this median since there are many more low-mass
than high-mass halos.  We can treat this with a simple approximation.
Using the assumed Gaussian distribution in $\log_{10} M_{\ast}$, assuming a
power-law halo mass function $dN/d\log_{10} M_h \propto M_h^{-\gamma}$ with
$\gamma \approx 0.9$ \citep[e.g.][]{tinker08}, and taking a local
power-law relationship $M_{\ast} \propto M_h^\beta$, we find an offset
of $-\gamma \ln(10) \sigma_{\ast}^2 \beta^{-2}$.  The distribution
will be a Gaussian in $\log_{10} M_\ast$ with dispersion $\beta^{-1}
\sigma_{\ast}$.

We find the final distributions in $\log_{10} M_h$ have
means and dispersions $(12.55,0.25)$ for the data given in
\citet{yang09}, and $(12.57,0.40)$ for the data given in
\citet{behroozi10}.  Uncertainty in the stellar mass adds
to the intrinsic dispersion,
implying the uncertainty in the halo mass may be even larger than
given here.  Our best value of $\log_{10} M_h = 12.33 \pm 0.10$
sits comfortably within these distributions.

Our detailed fitting of the GSS thus alleviates the tension between
M31's halo mass and the general galaxy population.  Of course, it is
too early to consider this issue definitively settled.  A host of
systematic effects not considered here may bias our result.  Ones that
may prove significant include dynamical friction, errors in the
assumed M31 distance, asphericity of the halo potential, and
deviations of the potential shape from our simple one-parameter
family.  However, our result does illustrate the potential of the
simulation sampling method.  The presence of several long streams in
the PAndAS survey \citep{alan09} suggests we can reduce both statistical
errors and degeneracies by applying constraints from several objects
simultaneously, which is quite feasible within a Bayesian framework.

\section{CONCLUSIONS} %7
\label{sec.conclusions}
We have combined N-body models of the Giant Southern Stream (GSS) and
related debris structures within M31 with MCMC sampling methods, to
describe our knowledge of the interesting physical parameters.  We
have added the sky pattern of RGB stars, as given by the INT
photometric survey of M31's halo, to the observations of GSS
positional and kinematic quantities used in previous stream fitting
work.  The combination of observables now tightly constrains the
model.

We find the stream's progenitor had a stellar mass at last pericentric
passage of $\log_{10} \, (M_s/\msun) = 9.5 \pm 0.1$, comparable to the LMC.
The time of this disruptive pericentric passage is tightly constrained
to $760 \pm 50 \Myr$.
Several lines of evidence suggest that the mass of dark matter in 
the progenitor was, at most, similar to the stellar mass.
We expect the debris from the progenitor's core
to be located in the NE Shelf.  Characteristic signatures in the
space of velocity versus radius may help localize the core debris 
in kinematic surveys.  
We find M31's virial mass is $\log_{10} \, M_{200} = 12.3 \pm 0.1$, 
alleviating the previous tension between observational virial mass
estimates and expectations from the general galactic population and
the timing argument.  

More generally, we expect the techniques used in this paper to be
useful in building informative models of other tidal debris
structures.  The combination of Bayesian methods with $N$-body
simulation requires significant investments of both computer time and
human effort.  Future work will surely yield technical advances on the
sampling techniques used here, and we have indicated some possible
directions for study.  However, even at the current level of
sophistication, the rewards of the method are significant, including
libraries of fairly sampled simulations and uncertainty estimates of
physical parameters that allow for scientifically meaningful
discussion.

\section*{ACKNOWLEDGMENTS}
We thank Tom Quinn and Joachim Stadel for the use
of PKDGRAV, and Josh Barnes for the use of ZENO.
%We also acknowledge a useful discussion with Anil Seth.
MAF and MDW acknowledge support by NSF grant AST-1009652 to 
the University of Massachusetts.
AB acknowledges support from NSERC through the Discovery Grant program.
PG acknowledges support by NSF grant AST-1010039
and NASA grant HST-GO-12055.
KG  acknowledges support from NASA through Hubble Fellowship
grant 51273.01 by the Space Telescope
Science Institute, which is operated by the Association of Universities 
for Research in Astronomy, Inc., for NASA, under contract NAS 5-26555.
We acknowledge useful conversations with 
Anil Seth, Elena d'Onghia, and John Dubinski.
%\input bibliography.tex
%how do I fix this?
\bibliographystyle{mn2e}
\bibliography{m31}
\label{lastpage}
\end{document}